\documentclass[a4paper,11pt]{article}
\pdfoutput=1
\usepackage{jheppub}
\bibstyle{JHEP}

\usepackage[T1]{fontenc}

\usepackage{enumitem}
\usepackage{empheq}
\usepackage{amsmath}
\usepackage{amssymb}
\usepackage{dsfont}
\usepackage{graphicx}
\usepackage{hyperref}
\usepackage{stackrel}
\usepackage{cancel}
\usepackage[utf8]{inputenc}  
\usepackage{tikz}
\usetikzlibrary{patterns,arrows,positioning,shapes.misc,decorations.pathmorphing}

\newcommand{\bR}{\mathbb{R}}

\newcommand{\cC}{\mathcal{C}}
\newcommand{\cD}{\mathcal{D}}

\newcommand{\cK}{\mathcal{K}}
\newcommand{\cM}{\mathcal{M}}

\newcommand{\cO}{\mathcal{O}}
\newcommand{\cR}{\mathcal{R}}
\newcommand{\cV}{\mathcal{V}}
\newcommand{\cW}{\mathcal{W}}

\newcommand{\vol}{\mathrm{vol}}

\newcommand{\ads}{\text{AdS}}
\newcommand{\cft}{\text{CFT}}
\newcommand{\lcft}{l_\text{CFT}}

\DeclareMathOperator{\arcosh}{arcosh}

\newcommand\figblue{gray!30!blue!30!white}
\newcommand\figyellow{yellow!40!white}

\newcommand{\be}    {\begin{equation}}
\newcommand{\ee}    {\end{equation}}

\newcommand{\req}[1]	{(\ref{#1})}
\newcommand{\figref}[1]	{{Fig.~\ref{#1}}}
\newcommand{\secref}[1]	{{Sec.~\ref{#1}}}

\begin{document}
%%%%%%%%%%%%%%%%%%%%%%%%%%%%%%%%%%%%%%%%%%%%%%%%%%%%%%%%%%%%%%%%%%%%%%%%%%%%%%%%

\title{Holographic Subregion Complexity from\\Kinematic Space}

\author[a]{Raimond Abt,}

\author[a]{Johanna Erdmenger,}

\author[a]{Marius Gerbershagen,}

\author[a]{\mbox{Charles M. Melby--Thompson},}

\author[a]{Christian Northe}

\affiliation[a]{Institut f{\"u}r Theoretische Physik und Astrophysik,\\Julius-Maximilians-Universit{\"a}t W{\"u}rzburg, Am Hubland, 97074 W\"urzburg, Germany}

\abstract{%
We consider the computation of volumes contained in a spatial slice of $\ads_3$ in terms of observables in a dual CFT. 
Our main tool is kinematic space, defined either from the bulk perspective as the space of oriented bulk geodesics, or from the CFT perspective as the space of entangling intervals.
We give an explicit formula for the volume of a general region in a spatial slice of $\ads_3$ as an integral over kinematic space. 
For the region lying below a geodesic, we show how to write this volume purely in terms of entangling entropies in the dual CFT.
This expression is perhaps most interesting in light of the complexity $\!=\!$ volume proposal, which posits that complexity of holographic quantum states is computed by bulk volumes. 
An extension of this idea proposes that the holographic subregion complexity of an interval, defined as the volume under its Ryu-Takayanagi surface, is a measure of the complexity of the corresponding reduced density matrix.
If this is true, our results give an explicit relationship between entanglement and subregion complexity in CFT, at least in the vacuum. 
We further extend many of our results to conical defect and BTZ black hole geometries.}


\maketitle

%%%%%%%%%%%%%%%%%%%%%%%%%%%%%%%%%%%%%%%%%%%%%%%%%%%%%%%%%%%%%%%%%%%%%%%%%%%%%%%%
\section{Introduction}
\label{sec: Introduction}
%%%%%%%%%%%%%%%%%%%%%%%%%%%%%%%%%%%%%%%%%%%%%%%%%%%%%%%%%%%%%%%%%%%%%%%%%%%%%%%%
A central component of the AdS/CFT duality is the question of how information in the boundary CFT is encoded in its dual gravitational theory. 
One lesson learned in the past decade is that there appear to be deep connections between quantum information-theoretic objects on the boundary and geometric quantities in the bulk.
The prototypical example of such a relation is the Ryu-Takayanagi (RT) proposal \cite{2006PhRvL..96r1602R}, 
which states that the entanglement entropy of a subregion $A$ on a given 
constant time slice on the CFT side is given (at leading order in $1/N$) by the area of an extremal codimension two bulk surface with the same boundary as $A$, known as the Ryu-Takayanagi surface. 
Other proposals relating quantum information to geometry include entanglement of purification \cite{Takayanagi:2017knl,Nguyen:2017yqw}, Fisher information \cite{Banerjee:2017qti, Lashkari:2015hha}, and fidelity susceptibility \cite{MIyaji:2015mia,Alishahiha:2017cuk,Gan:2017qkz,Flory:2017ftd}.
Quantum error correcting codes, describing quantum states that are maximally protected against erasure, can be constructed from tensor networks inspired by, and intended to mimic the properties of, holographic boundary states \cite{Pastawski:2015qua, Pastawski:2016qrs}. 
The relation between quantum information and geometry even allowed the derivation in \cite{Lashkari:2013koa} of the linearized Einstein's equations about AdS using entanglement entropy. 

The information-theoretic quantity we are interested in here is \emph{complexity}~\cite{Papadimitrou}. 
The complexity of a pure state within quantum information is computed with respect to a reference state and a set of basic unitary operators, called \textit{gates}. 
It is defined to be the minimum number of gates that must be applied to the reference state in order to map it to the desired state.%
\footnote{When working with a discrete set of unitary gates, we actually demand that the state is reproduced up to a given accuracy. A cleaner definition can be made by taking the limit as the gates approach the identity, which corresponds to introducing a measure on paths in $U(N)$ \cite{Nielsen:2005}.}
For quantum theories defined on a finite-dimensional Hilbert space (e.g. spin systems), this definition is straightforward. 
The definition of the complexity of a state in field theory, however, is much less clear-cut, although there has been some progress for free field theories \cite{Chapman:2017rqy, Jefferson:2017sdb, Hackl:2018ptj, Khan:2018rzm}. 

\begin{figure}[t]
\begin{center}
\includegraphics[scale=0.15]{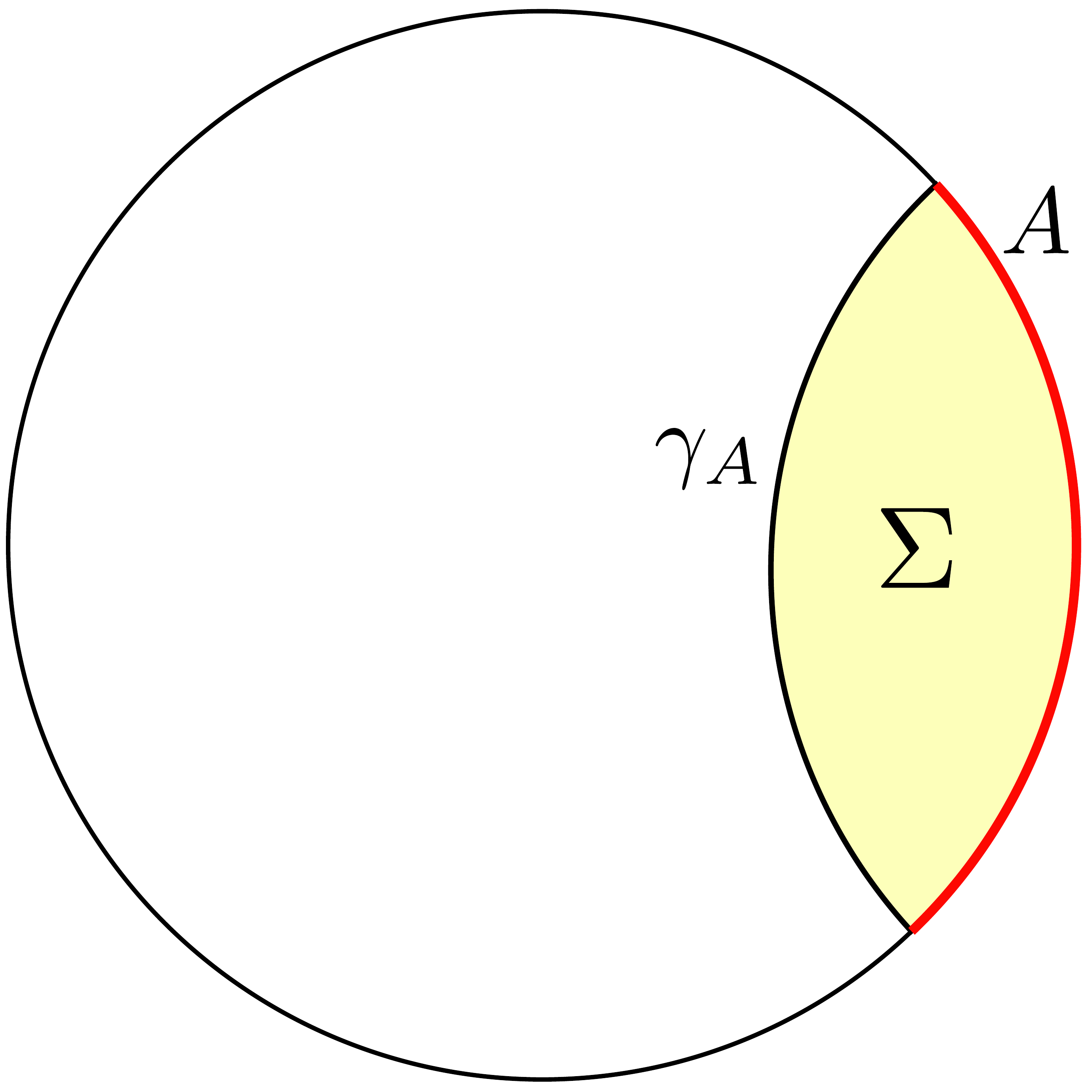} 
\end{center}
\caption{In AdS$_3$/CFT$_2$ the RT proposal states that the
			entanglement entropy of the region $A$ is given by the length of
			the geodesic $\gamma_A$ in the constant time slice that connects
			the boundary points of $A$. The volume of the region $\Sigma$
			below $\gamma_A$ is proposed to be a measure for the complexity of the reduced density matrix corresponding to $A$.}
\label{fig: Sigma}
\end{figure}

Computing the complexity of a state using the holographic correspondence has recently garnered a good deal of attention. 
The first proposals for bulk geometric quantities dual to field theory complexity posited that the complexity of the evolution of two copies of a CFT initially entangled in the thermofield double state is dual either to the volume of the Einstein-Rosen bridge \cite{Susskind:2014rva}, or the action of a Wheeler-DeWitt patch \cite{BrownSusskind}, of the two-sided black hole geometry.
The first proposal relates complexity to a bulk volume, and is known as the `complexity equals volume' (complexity $\!=\!$ volume) proposal.
Based on complexity $\!=\!$ volume, Alishahiha proposed \cite{Alishahiha:2015rta} that, for a given entangling region $A$, the volume enclosed by the corresponding RT surface computes the complexity of the reduced density matrix of $A$ (see \figref{fig: Sigma} for the case of $\ads_3/\cft_2$). 
This object, called the \emph{(holographic) subregion complexity}, has been computed in a number of cases, e.g. \cite{Ben-Ami,Carmi:2017jqz,Carmi:2017ezk,Chen:2018mcc}.

We cannot as yet study subregion complexity as an entry in the AdS/CFT dictionary, however, because it is not clear how to define it in field theory. 
Since it is associated to the reduced density matrix, a possible approach would be to use recent work on mixed state complexities \cite{Agon:2018zso}. 
This work associates to any mixed state $\rho$ two basic measures of complexity: the \emph{spectrum complexity}, which measures the difficulty of constructing a mixed state $\rho_{spec}$ with the same spectrum as $\rho$, and \emph{basis complexity}, measuring the difficulty of constructing $\rho$ from $\rho_{spec}$. 
Applying these constructions to reduced density matrices may yield a natural field theory definition of subregion complexity. 
Comparing the properties of a complexity so defined to holographic subregion complexity is an interesting problem for future work, but lies beyond the scope of the present paper.
Instead, we will take a different approach to this issue. 

The aim of this paper is to define a quantity in $\cft_2$ that, when applied to a CFT with a weakly curved holographic dual, reproduces the holographic subregion complexity of vacuum $\ads_3$. 
We present and prove a formula for the volume of an arbitrary spatial bulk region as the integral over all geodesics of the lengths of geodesic segments intersecting that region. 
We refer to this as the \textit{volume formula}. 
We apply the volume formula to the region under the RT surface to obtain an integral expression for holographic subregion complexity written purely in terms of CFT entanglement entropy. 
This integral expression defines a field theory quantity in any CFT, which reduces to the holographic subregion complexity for CFTs with weakly curved holographic duals.%
\footnote{We gave preliminary results in this direction previously in \cite{Abt:2017pmf}.}
We then extend our results to non-vacuum states whose dual geometries are described by quotients of vacuum $\ads_3$: conical defects, dual to primary states, and black holes, dual to a system at finite temperature.

\medskip
Our approach is based on the \emph{kinematic space} formalism \cite{Czech}. 
The kinematic space of a time slice of a bulk geometry is the space of oriented bulk geodesics in that slice%
\footnote{Later generalizations \cite{Czech:2016xec,deBoer:2016pqk} drop the restriction to constant time slices, but these are not relevant to this work.} 
ending on the boundary \cite{Czech:2014ppa,Czech,Czech:2015kbp,Czech:2017zfq}.
Kinematic space is therefore parametrized by pairs of points in a time slice of the dual CFT.
%
%-------------------------------------------
%-------------------------------------------
%
It was shown in \cite{Czech:2014ppa} that it is equipped with a natural volume form $\omega$, constructed from derivatives of the (regularized) length of geodesics.
The length of a bulk geodesic $\gamma_A$ sharing its boundary points with a CFT interval $A$ (\figref{fig: Sigma}) corresponds, by the RT formula, to the entanglement entropy $S(A)$ of $A$. 
In this way, we can alternatively identify kinematic space as the space of CFT intervals, and the length of $\gamma_A$ with $S(A)$.
Hence we may treat kinematic space as a geometry associated to any $\cft_2$, by defining it as the space of all CFT intervals and equipping it with the volume form $\omega$. 
This construction makes no reference to any holographic dual geometry.
It was shown in \cite{Czech:2014ppa,Czech,Czech:2015kbp} that detailed properties of the bulk geometry --- specifically, the perimeter of an arbitrary bulk curve --- can be computed as the integral with respect to the measure $\omega$ of the intersection number of geodesics with the curve. 
Applied to an AdS/CFT dual pair, this relates the entanglement pattern of the CFT to the geometry of its gravitational dual. 

In the same way, we may express the volume formula --- a bulk geometric relation --- as a computation purely within CFT.
We stress that for holographic subregion complexity, this expression requires no reference to a bulk geometry; 
for the vacuum state we obtain an expression for the holographic subregion complexity in terms of entanglement entropies alone. 
The bulk volumes that are associated with holographic subregion complexity are, however, divergent, and must be regularized. 
The cutoff scheme for the volume formula considered in our first presentation of this method \cite{Abt:2017pmf} corresponds to a radial cutoff in the bulk, and is rather opaque from the CFT perspective. 
We offer an alternative cutoff scheme that is more natural from the field theory and kinematic space points of view. 

The volume formula applies not just to vacuum $\ads$, but can also be adapted to its quotients. 
We give its explicit form in conical defect and static BTZ black hole geometries. 
Using the volume formula to compute holographic subregion complexity for these geometries, we confirm that it reproduces the results of direct calculations on the gravity side. 
An important proviso, however, is that for the conical defect and BTZ black hole geometry, boundary intervals and oriented geodesics are no longer in one-to-one correspondence.
Thus, knowledge of single interval entanglement entropies alone is not enough to compute the holographic subregion complexity in these states.

\medskip
The paper is organized as follows. 
Section \ref{sec: review of kinematic space} reviews those aspects of kinematic space relevant to this work. 
We present and prove the volume formula in section \ref{sec: volume formula proof}. 
In section \ref{sec: vacuum subregion complexity} we apply this formula to subregion complexity in the vacuum. 
We present an expression for it in terms of entanglement entropies and use it to compute subregion complexity in global AdS$_3$ and the Poincar\'e patch. 
In section \ref{sec: excited states} we consider primary states (dual to conical defects) and thermal states (dual to BTZ black holes). 
Here we introduce the appropriate kinematic spaces, and discuss the corresponding volume formulae. 
We discuss possible interpretations of the volume formula and present our conclusions in section \ref{sec: discussion}.

%%%%%%%%%%%%%%%%%%%%%%%%%%%%%%%%%%%%%%%%%%%%%%%%%%%%%%%%%%%%%%%%%%%%%%%%%%%%%%%%
\section{Review of Kinematic Space}
\label{sec: review of kinematic space}
%%%%%%%%%%%%%%%%%%%%%%%%%%%%%%%%%%%%%%%%%%%%%%%%%%%%%%%%%%%%%%%%%%%%%%%%%%%%%%%%
The concept of \emph{kinematic space} as a tool for studying the AdS/CFT correspondence was introduced in \cite{Czech:2014ppa}.
The utility of the kinematic space formalism lies in its ability to explicitly decode parts of the correspondence between bulk geometry and boundary information, making it an ideal starting point for studying bulk volumes in terms of CFT. 
In this section we review the basic concepts of kinematic space required for this work.

The RT formula suggests a strong relationship between entanglement and geometry, but does not immediately tell us how to construct the bulk geometry. 
One of the first steps toward making this correspondence more precise was the result of \cite{Balasubramanian:2013lsa} that the perimeter of a closed bulk curve could be constructed from derivatives of the entanglement entropy in terms of a quantity called \emph{differential entropy}. 
Kinematic space provides a natural framework for these concepts \cite{Czech:2014ppa,Czech,Czech:2015kbp}. 
It was also noted there that, in the special case of vacuum $\ads_3$, the perimeter formula reduces to known results from the field of \emph{integral geometry} (see e.g. \cite{Santalo}).

Consider an asymptotically AdS$_3$ spacetime $\cM$, i.e., a spacetime whose asymptotic behavior matches that of $\ads_3$:
\begin{equation}\label{eq: AdS}
	ds^2
	\sim
	-\frac{r^2}{L^2}dt^2+L^2\frac{dr^2}{r^2}+r^2d\phi^2
	\qquad
	\text{as}
	\quad
	r\to \infty\,,
\end{equation}
where $\phi\sim\phi+2\pi$ and $L$ is the AdS radius.%
\footnote{Kinematic space is constructed analogously for other geometries, such as the Poincar\'e patch, which we discuss later.}
We consider only the case where $\cM$ is static, with Killing time $t$.

\begin{figure}
\begin{center}
\includegraphics[scale=0.2]{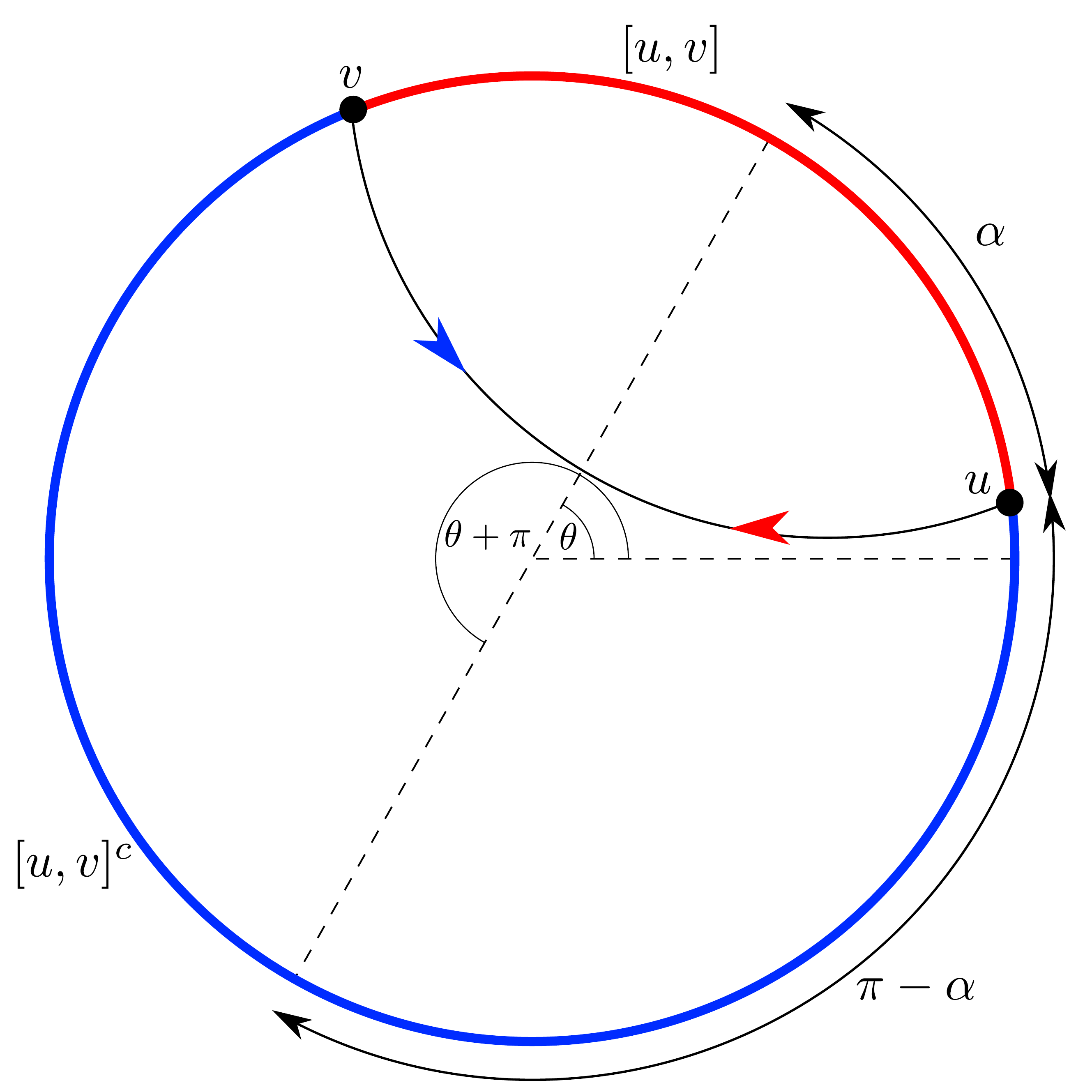} 
\end{center}
\caption{We can parametrize geodesics via
			their endpoints $u$ and $v$ or via the position of their center
			$\theta$ and their opening angle $\alpha$. The tuples 
			$(\theta,\alpha)$ and $(\theta+\pi, \pi-\alpha)$ correspond
			to the same geodesic, but with opposite orientation.			
			The geodesic with the orientation of the red arrow is associated with the
			entangling interval $[u,v]$, the geodesic with the orientation
			of the blue arrow is associated with the complement $[u,v]^c$.}
\label{fig:parametrization geodesics}
\end{figure}

\begin{figure}
\begin{center}
\includegraphics[scale=0.22]{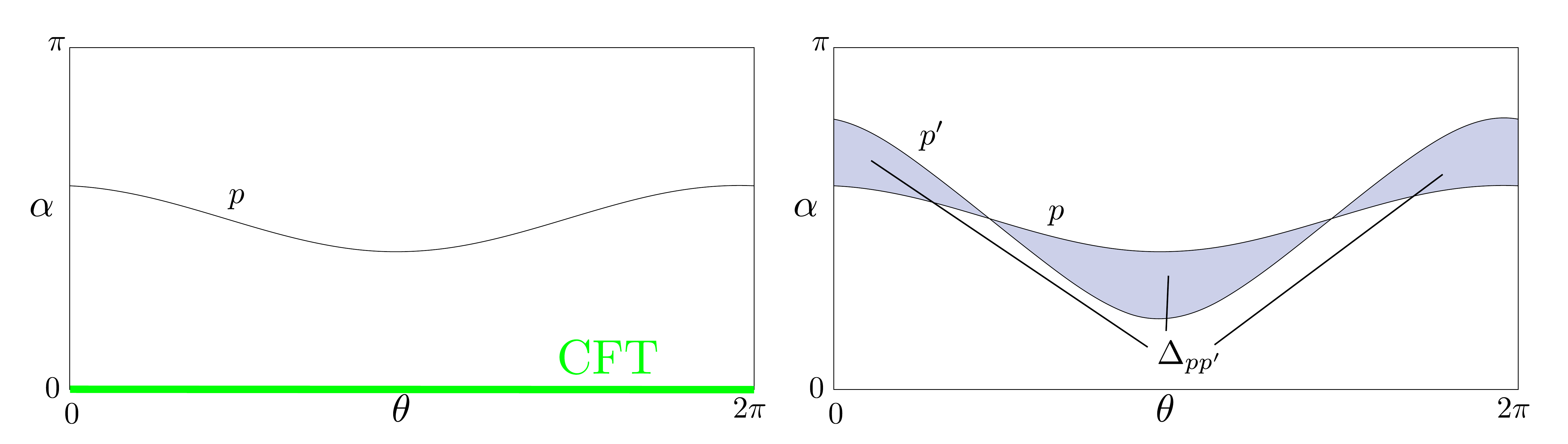} 
\end{center}
\caption{A point $p$ that lies in the constant time slice of asymptotic
			AdS$_3$ is associated with the set of all geodesics
			that intersect $p$ (LHS). This set is a curve in $\cK$,
			the so-called point
			curve of $p$. The geodesic distance of two points $p$ and $p'$ is
			given, up to a proportionality factor, by the volume of the region
			$\Delta_{pp'}$ in $\cK$ that is bounded by the point curves of $p$
			and $p'$ (RHS). Since $(\theta, \alpha=0)$ correspond to boundary
			points of $\ads_3$, the lower boundary of $\cK$ is identified with the constant time 
			slice of the CFT depicted in green (LHS).
			}
\label{fig:point curves}
\end{figure}

Fix a spatial slice given by $t=\text{constant}$.
Its kinematic space $\cK$ is the space of all oriented boundary-anchored geodesics that lie inside the slice. 
For simplicity, we assume that for any given pair of boundary points $u$, $v$ there is a unique oriented geodesic running from $u$ to $v$. 
This uniqueness is guaranteed in particular for geometries sufficiently close to pure AdS$_3$.
A point in $\cK$ -- that is, a geodesic -- is specified by the location $\phi=u,v$ of its endpoints, making $(u,v)$ a coordinate system on $\cK$.
An alternative parametrization uses the midpoint $\theta$ of the interval $[u,v]$ together with its opening angle $\alpha$ (\figref{fig:parametrization geodesics}),
\begin{equation}
	u = \theta - \alpha\,,
	\quad
	v = \theta + \alpha\,.
\end{equation}
%-------------------------------------------
%-------------------------------------------
%
As depicted in \figref{fig:parametrization geodesics}, $(\theta+\pi,\pi-\alpha)$ encodes the same geodesic as $(\theta,\alpha)$, but with the opposite orientation.
A bulk point $p$ is encoded in kinematic space as the set of all geodesics containing $p$. 
This set is a curve in $\cK$, the so-called \textit{point curve} (see \figref{fig:point curves}).
%
%-------------------------------------------
%-------------------------------------------

Given our assumptions, the geodesics $(u,v)$ are in one-to-one correspondence with the intervals $[u,v]$, so we may
interpret $\cK$ as the space of entangling regions of the CFT and consider the entanglement entropy $S(u,v)$ to be a function on it.
For a holographic CFT, this quantity is given at leading order in $1/N$ by the Ryu-Takayanagi formula:
\begin{equation}
\label{eq: RT}
	S(u,v)
	=
	\frac{\ell(u,v)}{4 G_N}\,.
\end{equation}
Here $\ell(u,v)$ denotes the length of the geodesic $(u,v)$, regularized for example by truncating at a large but finite value of $r$, and $G_N$ is the bulk Newton's constant.
The key observation of~\cite{Czech} was that $S$ induces a natural metric $ds^2_\cK$ on $\cK$, together with the corresponding volume form $\omega$:
\begin{align}
	\label{eq: ds kin. sp.}
	ds^2_\cK &= \partial_u\partial_vS\,du\,dv
	= \frac{1}{2}(\partial^2_\theta-\partial^2_\alpha)S\,(-d\alpha^2+d\theta^2) \,,
	\\
	\label{eq: omega}
	\omega
	&= \partial_u\partial_v S\,du\wedge dv
	= \frac{1}{2}(\partial^2_\theta-\partial^2_\alpha)S\,d\theta\wedge d\alpha\,.
\end{align}
In integral geometry the volume form is known as the \textit{Crofton form}.
In the following sections we will only consider geometries that are invariant under translations, meaning $S$ depends only on the length $v-u=2\alpha$ of the entangling interval and not its particular position. 
In this situation, \eqref{eq: ds kin. sp.} and \eqref{eq: omega} simplify to
\begin{equation}
\label{eq: ds and omega for symmetric spaces}
	ds^2_\cK
	=
	-\frac{1}{2}\partial^2_\alpha S(-d\alpha^2+d\theta^2)\,,
	\qquad
	\omega
	=
	-\frac{1}{2}\partial^2_\alpha Sd\theta\wedge d\alpha\,.
\end{equation}
The metric $ds^2_{\cK}$ is Lorentzian, and $u$ and $v$ are light-cone coordinates.

The geometric structure (\ref{eq: ds kin. sp.}, \ref{eq: omega}) of $\cK$ encodes useful information about the bulk geometry. 
For example, in pure $\ads_3$ point curves are known to be spacelike geodesics on $\cK$ \cite{Czech:2014ppa}.%
\footnote{In \cite{Czech:2014ppa} it was shown that point curves are geodesics for several geometries, such as global AdS$_3$, conical defects and BTZ black holes.} 
Furthermore it is possible to express the geodesic distance $d(p,p')$ between two bulk points $p$ and $p'$ as an integral in kinematic space \cite{Czech}:
\begin{equation}
\label{eq: geodesic distance}
	\frac{d(p,p')}{4G_N}
	=
	\frac{1}{4}\int_{\Delta_{pp'}}\omega\,.
\end{equation}
Here $\Delta_{pp'}\subset\cK$ is the set of all geodesics separating $p$ and $p'$. 
$\Delta_{pp'}$ turns out to be the region bounded by the point curves of $p$ and $p'$, as depicted in \figref{fig:point curves}.

From here on $(\theta,\alpha)$ will also denote entangling intervals and we will view $\cK$ as the space of these.
In this picture we can understand the causal structure of $\cK$ in an intuitive way:
$(u_1,v_1)$ lies in the past of $(u_2,v_2)$ if $[u_1,v_1]\subset[u_2,v_2]$.
Note that the orientation reversal $(\theta+\pi,\pi-\alpha)$ of the geodesic $(\theta,\alpha)$ is spacelike related to it; this is because
$(\theta+\pi,\pi-\alpha)$ corresponds to the complement of the entangling interval $(\theta,\alpha)$, as seen in \figref{fig:parametrization geodesics}.
The interpretation of $\cK$ as the space of CFT intervals means that $\cK$ can be constructed for any CFT, regardless of the (non-)existence of a bulk dual.
Finally, as $\alpha\to 0$ the geodesic $(\alpha,\theta)$ collapses to the boundary point $\phi=\theta$.
Therefore, the lower boundary $\cK$, $\alpha=0$, can be identified with the CFT circle (see \figref{fig:point curves}).
This observation plays an important role in later sections.

It is the RT proposal that connects kinematic space to quantum information.
Equation \eqref{eq: RT} tells us that $\ell(u,v)$ computes the entanglement entropy of the interval $[u,v]$. 
This connection allows one to express bulk lengths, as in \eqref{eq: geodesic distance}, and volumes, which we study in \secref{sec: volume formula proof}, as integrals over derivatives of entanglement entropies.
In this way, the information-theoretic properties of a constant time slice in the CFT encode the geometry of the corresponding constant time slice in the bulk.

In particular, the Crofton form $\omega$ can be interpreted as an infinitesimal version of the \textit{conditional mutual information} of two intervals $A$ and $B$ with respect to a third interval $C$,
\begin{equation}
	I(A,B|C)
	=
	S(AC)
	+
	S(BC)
	-S(ABC)
	-S(C)\,.
\end{equation}
We recover the Crofton form from the infinitesimal conditional mutual information of the neighboring intervals $A=[u-du,u]$, $B=[v,v+dv]$, $C=[u,v]$ ~\cite{Czech}:
\begin{equation}
	I(A,B|C)
	\approx
	\partial_u\partial_v S\,du\,dv\propto\omega\,.
\end{equation}
Note that we can also motivate the causal structure of $\cK$ by requiring $(u_1,v_1)$ to lie in the past of $(u_2,v_2)$ if $[u_1,v_1]\subset[u_2,v_2]$.
This immediately leads to
\begin{equation}
	ds_\cK^2\propto du\,dv\,.
\end{equation}
The proportionality factor, $\partial_u\partial_vS$, is fixed by demanding that the volume form match the Crofton form.
Consequently, the geometry of $\cK$ can be constructed from the CFT side without reference to the bulk. 
This will be important for us when we construct a field theory expression for subregion complexity.

%%%%%%%%%%%%%%%%%%%%%%%%%%%%%%%%%%%%%%%%%%%%%%%%%%%%%%%%%%%%%%%%%%%%%%%%%%%%%%%%
\section{The Volume Formula}
\label{sec: volume formula proof}
\begin{figure}
\begin{center}
\includegraphics[scale=0.15]{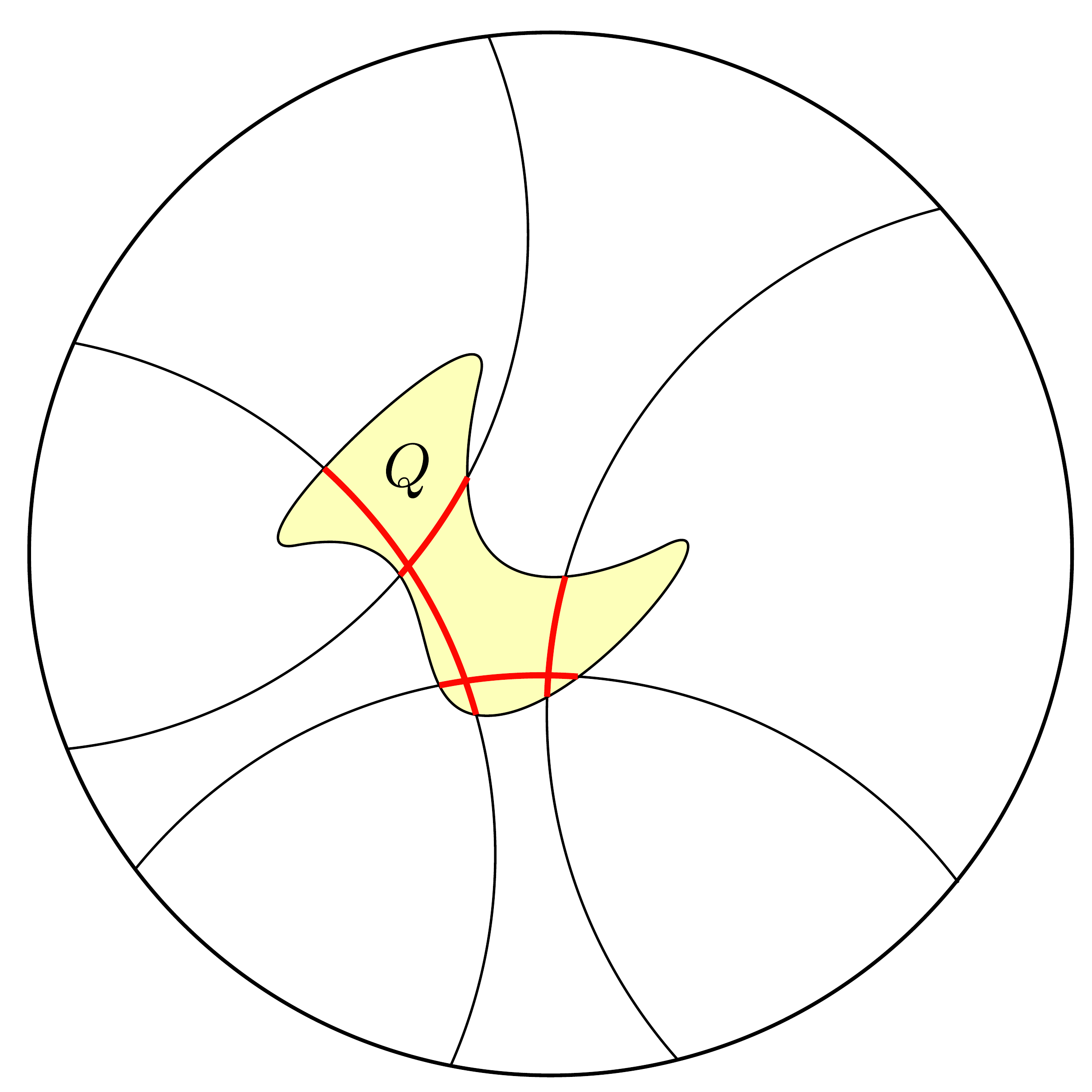} 
\end{center}
\caption{The volume of a region $Q$ on the constant time slice is given by
			an integral over the chord lengths of all geodesics. The chord
			length of a geodesic is the length of the segment of the geodesic
			that lies inside of $Q$ (depicted in red).}
\label{fig:arbit Q}
\end{figure}

%%%%%%%%%%%%%%%%%%%%%%%%%%%%%%%%%%%%%%%%%%%%%%%%%%%%%%%%%%%%%%%%%%%%%%%%%%%%%%%%
The goal of this paper is to establish and apply the following formula for the volume of a bulk region $Q$ as a kinematic space integral,
\begin{equation}
\label{eq:volume formula}
	\frac{\vol(Q)}{4G_N}
	=
	\frac{1}{2\pi}\int_\cK \lambda_Q\omega\,,
\end{equation}
which we first presented in \cite{Abt:2017pmf}.
Here $\lambda_Q(\theta,\alpha)$ is the \textit{chord length} of the geodesic
$(\theta,\alpha)$, defined to be the length of the intersection of the geodesic $(\theta,\alpha)$ with $Q$ (\figref{fig:arbit Q}). 
%
%-------------------------------------------
%-------------------------------------------
%
In the following sections we use it to derive an expression for holographic subregion complexity in the vacuum purely in terms of field theory quantities.

While formulae like this are known in integral geometry \cite{Santalo}, we present here a simple proof of \eqref{eq:volume formula} for the kinematic space of a constant time slice of global AdS$_3$ with metric
\begin{equation}
\label{eq: metric global AdS3}
	ds^2_{\ads_3}
	=
	-\frac{L^2+r^2}{L^2}dt^2+\frac{L^2}{L^2+r^2}dr^2+r^2d\phi^2\,.
\end{equation}
In this case the entanglement entropy is
\begin{equation}
\label{eq: EE for global AdS3}
 S(\alpha)
 =
 \frac{c}{3}\log\Bigl(\frac{2\lcft}{\epsilon}\sin(\alpha)\Bigr)\,,
\end{equation}
where $c=\frac{3L}{2G_N}$ is the central charge, $\lcft$ is the radius of the CFT circle and $\epsilon$ is the UV cutoff. 
The corresponding metric and Crofton form are 
\begin{equation}
	ds^2_{\cK} = \frac{c}{6}\frac{1}{{\sin^2}\alpha}(-d\alpha^2+d\theta^2) \,,
	\qquad
	\omega=\frac{c}{6}\frac{1}{{\sin^2}\alpha}d\theta\wedge d\alpha\,.
\label{eq:vacuum kinematic space}
\end{equation}

The strategy we pursue begins by verifying the volume formula for a disc $D_R$ of radius $R$ around the point $r=0$ in a constant time slice of AdS$_3$.%
\footnote{We presented this computation previously in \cite{Abt:2017pmf}.}
We next show that the integral in \eqref{eq:volume formula} shares with volumes certain characteristic properties such as non-negativity and additivity, and use these properties to extend the volume formula to annular arcs.
Using annular arcs it is possible to construct Riemann sums, which approximate the volume of $Q$ arbitrarily well, proving the volume formula in the limit.

\begin{figure}
\begin{center}
\includegraphics[scale=0.15]{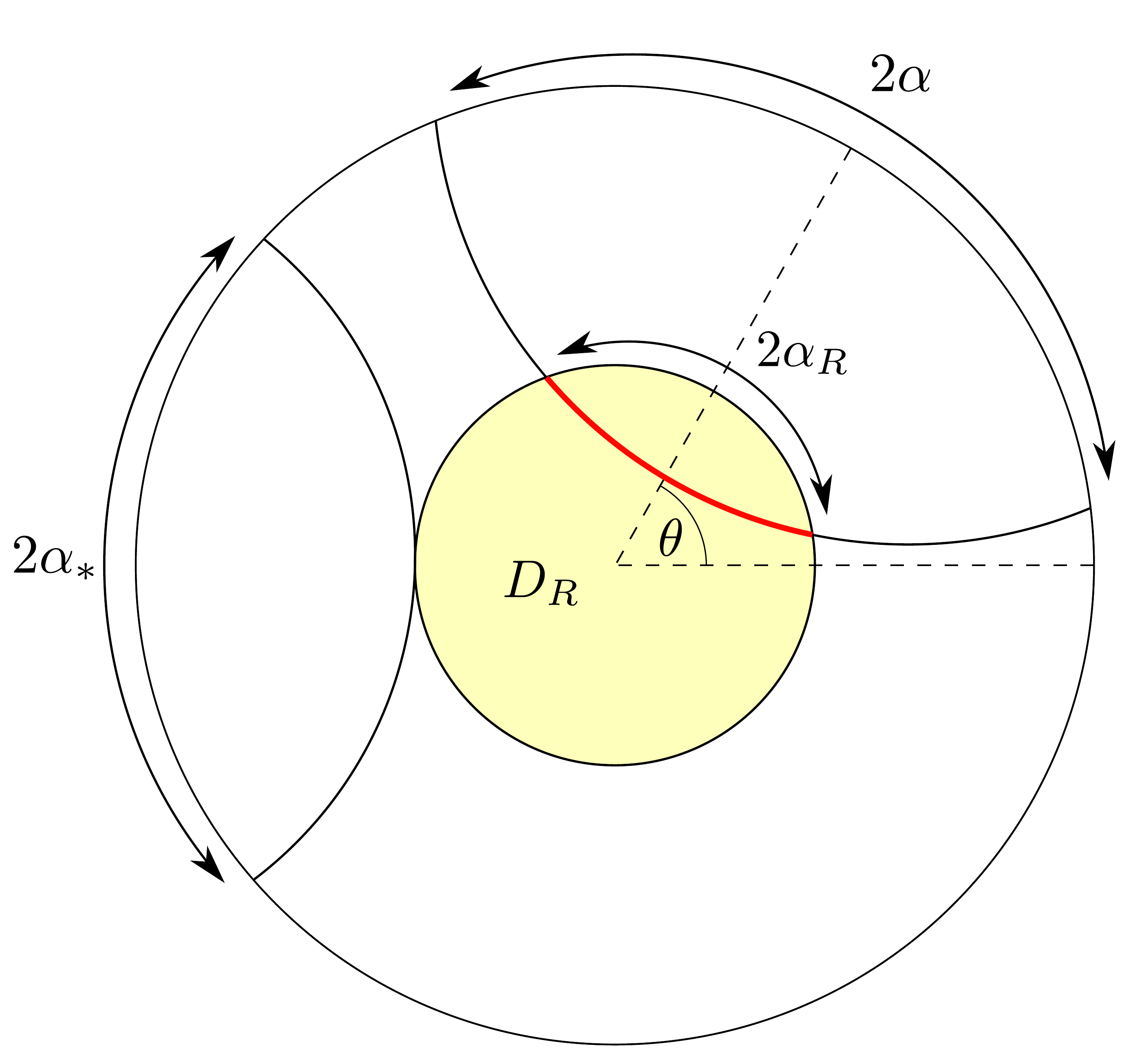} 
\end{center}
\caption{The disc $D_R$ associates an opening angle $\alpha_R$ on its boundary
			to each geodesic $(\theta,\alpha)$. Geodesics of the form
			$(\theta,\alpha_*)$ are tangent to $D_R$.}
\label{fig:Disc in AdS}
\end{figure}

Denoting the integral in \ref{eq:volume formula} by
\begin{equation}
	V(Q)
	\equiv
	\frac{2G_N}{\pi}\int_\cK \lambda_Q \omega \,,
\end{equation}
our proposal is that
\begin{equation}
\label{eq:V=vol}
	V(Q)
	=
	\vol(Q) \,.
\end{equation}
Let us first establish this for a disc $D_R$ ($r\le R$) of radius $R$. 
The chord length of the geodesic $(\theta,\alpha)$ for region $D_R$ is 
\begin{equation}
\label{eq:chord length for discs}
	   \lambda_{D_R}(\theta,\alpha)
	=
\begin{cases}
    L\arcosh(1+2\frac{R^2}{L^2}\sin^2(\alpha_R))\,,& \text{if } \alpha_*\leq\alpha\leq \pi-\alpha_*\\
    0\,,              & \text{otherwise.}
\end{cases}\,
\end{equation}
Here $\alpha_R$ is the opening angle of the geodesic $(\theta,\alpha)$ on the boundary of $D_R$ (\figref{fig:Disc in AdS}), and satisfies
\begin{equation}
	\frac{R}{\sqrt{L^2+R^2}}\cos(\alpha_R)
	=
	\cos(\alpha)\,.
\end{equation}
The angle $\alpha_*$ is given by
\begin{equation}
	\cos(\alpha_*)
	=
	\frac{R}{\sqrt{L^2+R^2}} \,,
\end{equation}
and specifies the family of geodesics $(\theta,\alpha_*)$ tangent to $D_R$ (\figref{fig:Disc in AdS}).
%
%-------------------------------------------
%-------------------------------------------
%
Since $\lambda_{D_R}$ vanishes for
$\alpha\not\in [\alpha_*,\pi-\alpha_*]$ (see \eqref{eq:chord length for discs}),
$V(D_R)$ takes the form
\begin{equation}
	V(D_R)
	=
	-\frac{1}{4\pi}
	\int_0^{2\pi}d\theta\int_{\alpha_*}^{\pi-\alpha_*}d\alpha \,
		\lambda_{D_R}\partial_\alpha^2\ell\,.
\end{equation}
By expressing $V(D_R)$ as an integral over $\alpha_R$ and integrating by parts, we find
\begin{equation}
\label{eq:disc volume}
\begin{split}
	V(D_R)
	&
	=
	\frac{1}{2}\int_0^\pi d\alpha_R(\partial_{\alpha_R}\lambda_{D_R})^2
	=
	\int_0^\pi d\alpha_R\frac{2L^2R^2\cos^2(\alpha_R)}{L^2+R^2\sin^2(\alpha_R)}
	\\
	&
	=
	2\pi L^2\Big(\sqrt{1+\frac{R^2}{L^2}}-1\Big)	
	\,,
\end{split}
\end{equation}
which is indeed the volume of the disc $D_R$.

Our next step is to establish the following important properties of $V$:
\begin{enumerate}[label=(\alph*)]
\item 
$V(Q)\ge 0$, with equality only when $Q=\emptyset$. 
This is simply due to the fact that it is the integral of a non-negative function with a positive volume form. 

\item
$V$ is additive,
\begin{equation}
\label{eq:additivity}
	V(Q\cup Q')
	=
	V(Q)+V(Q')-V(Q\cap Q') \,.
\end{equation}
Here, $Q$ and $Q'$ are any regions in the constant time slice of AdS$_3$.
This property is a direct consequence of the additivity of chord lengths,
\begin{equation}
\label{eq:additivity of lambda}
	\lambda_{Q\cup Q'}
	=
	\lambda_{Q}+\lambda_{Q'}-\lambda_{Q\cap Q'} \,.
\end{equation}

\item 
Non-negativity and additivity, together with $V(\emptyset)=0$, imply that $V$ is monotonic,
\begin{equation}
	V(Q)\leq V(Q')\quad\mbox{if }Q\subseteq Q'\,.
\end{equation}

\item 
$V$ is invariant under rotations around $r=0$.
This follows from the rotational invariance of the vacuum state (implying rotational invariance of the kinematic space measure) and of the chord length $\lambda_\Sigma$.

\end{enumerate}

\begin{figure}
\begin{center}
\includegraphics[scale=0.15]{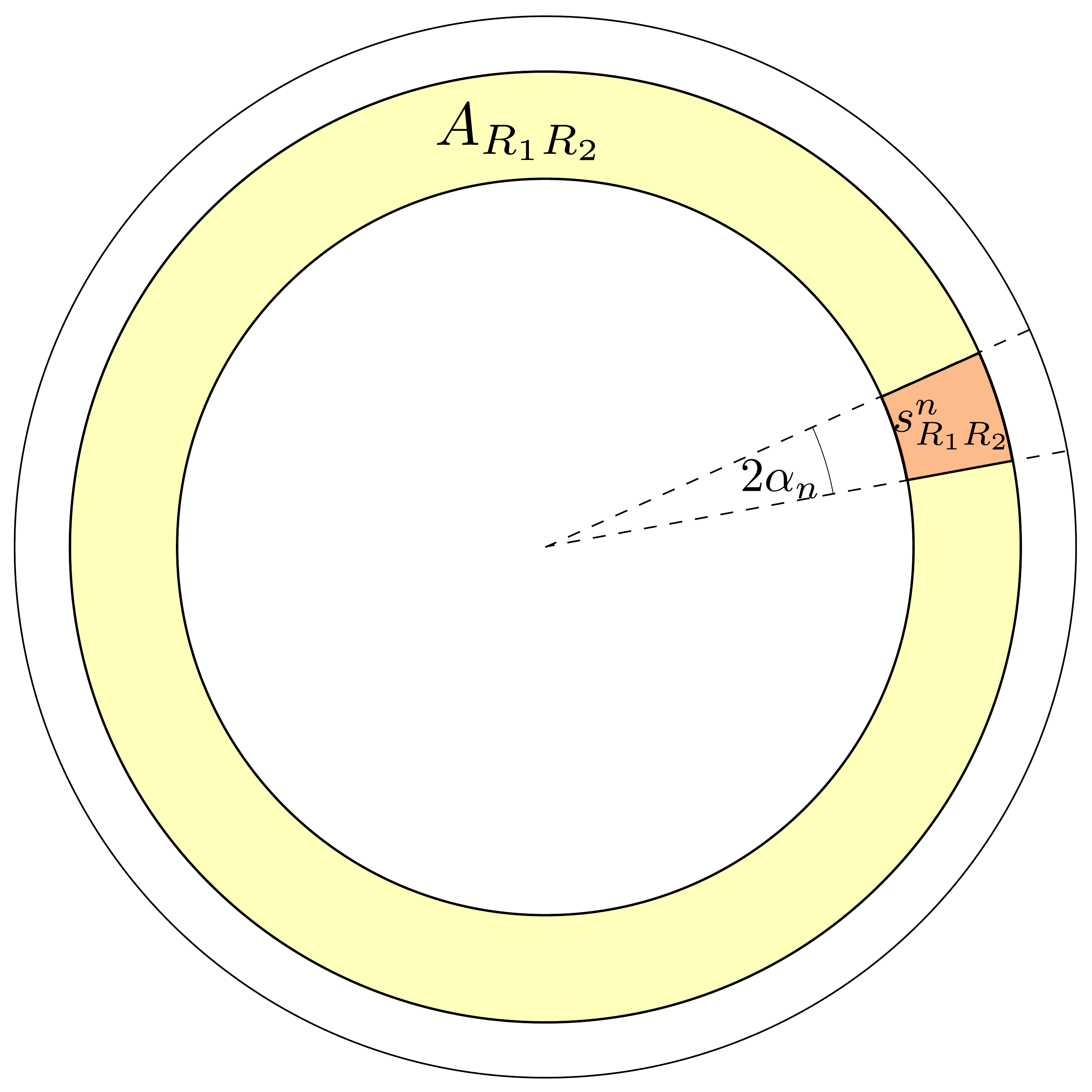} 
\end{center}
\caption{Annulus $A_{R_1R_2}$ of inner radius $R_1$ and outer radius
			$R_2$ and annulus segment $s^n_{R_1R_2}$ with opening angle
			$\alpha_n$.}
\label{fig:ring}
\end{figure}

We can prove \eqref{eq:V=vol} by taking advantage of properties (a)-(d).
Consider \eqref{eq:V=vol} for an annulus $A_{R_1R_2}$ of inner radius $R_1$ and outer radius $R_2$ centered around the origin (\figref{fig:ring}).
%
%-------------------------------------------
%-------------------------------------------
%
First note that, since the disc $D_{R_2}$ can be written as the union $D_{R_2}=D_{R_1}\cup A_{R_1R_2}$, additivity implies 
\begin{equation}
\label{eq:additivity for annulus}
	V(A_{R_1R_2})
	 =
  	V(D_{R_2})-V(D_{R_1}).
\end{equation}
We already know that the volume formula holds for $D_R$. 
Therefore, \eqref{eq:additivity for annulus} shows that it also holds for $A_{R_1R_2}$:
\begin{equation}
\label{eq: vol form for annuli}
	V(A_{R_1R_2})
	=
	\vol(D_{R_2})-\vol(D_{R_1})
	=
	\vol(A_{R_1R_2})\,.
\end{equation}
The second step is to verify the proposal for a segment
$s_{R_1R_2}^n$ of the annulus $A_{R_1R_2}$  (see \figref{fig:ring})
with opening angle
\begin{equation}
	\alpha_n
	\equiv
	\frac{\pi}{n}\,,
	\qquad
	n\in\mathbb{N} \,.
\end{equation}
Rotational invariance, additivity, and \eqref{eq: vol form for annuli} together imply
\begin{equation}
	V(s_{R_1R_2}^n)
	=
	\frac{1}{n}
	V(A_{R_1R_2})
	=
	\frac{1}{n}
	\vol(A_{R_1R_2})
	=
	\vol(s_{R_1R_2}^n)\,.
\end{equation}
So the proposal indeed holds for segments of annuli with opening angle
$\alpha_n$.

%-------------------------------------------
\begin{figure}
\begin{center}
\includegraphics[scale=0.15]{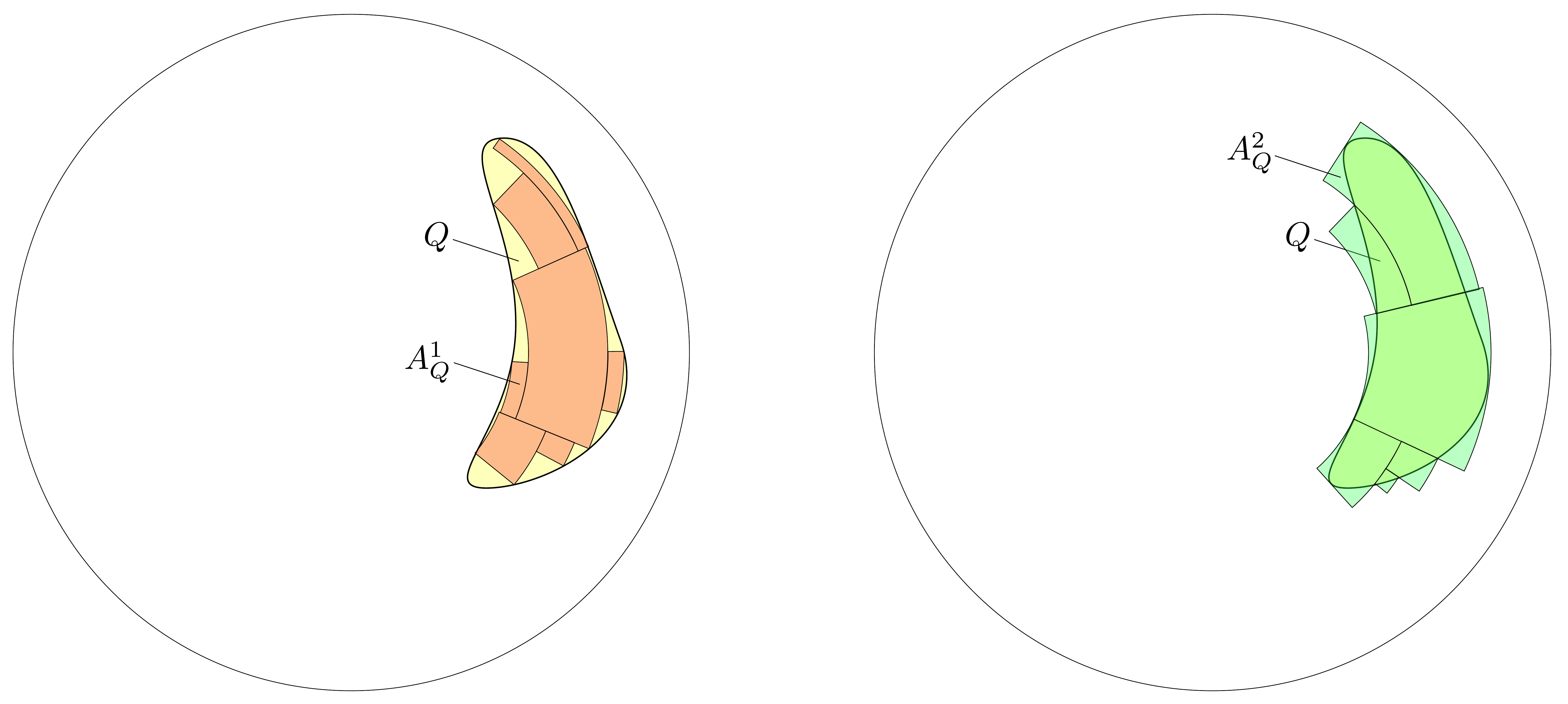} 
\end{center}
\caption{Approximation of an arbitrary set Q by annular arcs.
	The approximations $A^1_Q\subseteq Q$ and $A^2_Q\supseteq Q$ are depicted in red and green, respectively. 
}
\label{fig: approx Q}
\end{figure}
%-------------------------------------------

Now consider an arbitrary region $Q$. 
We can approximate $V(Q)$ arbitrarily well by approximating $Q$ by a disjoint union of sufficiently small annular arcs.
Examples of such approximations strictly contained in $Q$ (region $A_Q^1$) and strictly containing $Q$ (region $A_Q^2$) are shown in \figref{fig: approx Q}.
Taking the limit where the arc size goes to zero proves the volume formula for arbitrary $Q$.

%%%%%%%%%%%%%%%%%%%%%%%%%%%%%%%%%%%%%%%%%%%%%%%%%%%%%%%%%%%%%%%%%%%%%%%%%%%%
\subsection*{Alternative Proof for the Poincar\'e patch}
%%%%%%%%%%%%%%%%%%%%%%%%%%%%%%%%%%%%%%%%%%%%%%%%%%%%%%%%%%%%%%%%%%%%%%%%%%%%
The volume formula~\eqref{eq:volume formula} also holds in Poincar\'e patch coordinates. 
We offer here a proof of the Poincar\'e patch volume formula by a direct computation of the volume of an infinitesimally thin rectangle.

The metric of a constant time slice of the Poincar\'e patch is 
\begin{equation}
  ds^2 = L^2\frac{dx^2 + dz^2}{z^2} \,.
  \label{eq:poincare-patch-metric}
\end{equation}
The geodesics of this geometry are semicircles anchored at $z=0$ and are parametrized by two endpoints $u$ and $v$ at the boundary, or equivalently the circle's center $\chi$ and radius $\psi$.
Explicitly,
\be
\label{eq: kin sp coordinates for Pp}
z^2=\psi^2-(x-\chi)^2 \,, \qquad u = \chi-\psi\,,\;\; v = \chi+\psi \,.
\ee
Cutting off the geometry at $z=\epsilon$, the length of the geodesic is
\begin{equation}
\ell(u,v)= L\,\log\left(
	\frac{1+\sqrt{1-(\epsilon/\psi)^2}}
		{1-\sqrt{1-(\epsilon/\psi)^2}}\right) \,,
\label{eq:geodesic-length-poincare}
\end{equation}
which is of course translation-invariant. 
Applying (\ref{eq: ds kin. sp.}, \ref{eq: omega}) and sending $\epsilon\to 0$ gives the kinematic space volume form and metric
\begin{equation}
  \omega = \frac{c}{6} \frac{d\chi\wedge d\psi}{\psi^2} \,, 
  \qquad\quad
  ds_{\cK}^2 = \frac{c}{6} \frac{d\chi^2-d\psi^2}{\psi^2} \,.
\end{equation}
To simplify calculations we work with positively oriented geodesics ($\psi>0$), and multiply by two at the end.

To compute the volume of a bulk region $Q$, divide it into disjoint rectangles $R_i$ of finite height stretching from $z_{i,1}$ to $z_{i,2}$, but infinitesimal width $\delta x$.
Because the chord length is additive,
\be
\lambda_{Q} = \lambda_{\cup_i R_i} = \sum_i \lambda_{R_i} \,,
\ee
it suffices to show the volume formula for a general such rectangle $R$.
To first order in $\delta x$, only geodesics entering from the left and exiting from the right of $R$ contribute.
The length of this intersection is given by
\begin{equation}
\lambda_R = ds = \frac{L}{z}\sqrt{\delta x^2 +\delta z^2}
= \frac{L\, \delta x}{z^2}\sqrt{z^2 + (x-\chi)^2}
= L \frac{\psi\,\delta x}{\psi^2-(x-\chi)^2} 
\end{equation}
if $z_1<\psi^2-(x-\chi)^2<z_2$, and zero otherwise.
Setting $\psi_{1,2} = \sqrt{z_{1,2}^2+(x-\chi)^2}$, the integration region in kinematic space is now $\psi_1 < \psi < \psi_2$.
The volume formula then takes the form
\begin{equation}
\vol(R) = 2\cdot\frac{4G_N}{2\pi}
\int_{-\infty}^\infty d\chi
\int_{\psi_1}^{\psi_2}d\psi 
L \frac{\psi\,\delta x}{\psi^2-(x-\chi)^2}
\cdot \frac{c}{6}\frac{1}{\psi^2}
= L^2\delta x\Bigl(\frac{1}{z_1}-\frac{1}{z_2}\Bigr) \,,
\end{equation}
matching the volume computed directly from the Poincar\'e patch metric and proving the volume formula.

%%%%%%%%%%%%%%%%%%%%%%%%%%%%%%%%%%%%%%%%%%%%%%%%%%%%%%%%%%%%%%%%%%%%
\section{Vacuum Subregion Complexity}
\label{sec: vacuum subregion complexity}
%%%%%%%%%%%%%%%%%%%%%%%%%%%%%%%%%%%%%%%%%%%%%%%%%%%%%%%%%%%%%%%%%%%%
Having proved the volume formula \eqref{eq:volume formula}, we are in a position to derive an expression for subregion complexity in a vacuum state in terms of entanglement entropy. 
The holographic subregion complexity of a CFT interval was defined in \cite{Alishahiha:2015rta} to be $\frac{1}{8\pi G_NL}\vol(\Sigma)$, where $\Sigma$ is the region contained beneath its RT surface (\figref{fig: Sigma}).
Using the kinematic space parametrization of entangling intervals of section~\ref{sec: review of kinematic space}, we denote the boundary interval by $(\theta_\Sigma,\alpha_\Sigma)$.
The volume of $\Sigma$ is easily computed, either directly \cite{Alishahiha:2015rta,Ben-Ami,Carmi:2017jqz} or by making use of the Gauss-Bonnet theorem \cite{Abt:2017pmf}. 
In our companion paper \cite{Abt:2017pmf}, we defined the \emph{topological complexity} $\cC(\theta_\Sigma,\alpha_\Sigma)$ of the interval $(\theta_\Sigma,\alpha_\Sigma)$ to be given by the integral of the scalar curvature $\mathcal{R}$ of the constant time slice over $\Sigma$
\begin{equation}
\label{eq: topological complexity}
	\cC(\theta_\Sigma,\alpha_\Sigma)
	=
	-\frac{1}{2}\int_\Sigma d\sigma\,\mathcal{R}\,.
\end{equation}
The terminology reflects its connection to the Gauss-Bonnet theorem.
In this paper we only consider geometries with constant $\mathcal{R}$, in which case our definition \eqref{eq: topological complexity} is proportional to the volume, 
\begin{equation}
\label{eq:subregion complexity}
	\cC(\theta_\Sigma,\alpha_\Sigma)
	=
	-\frac{\mathcal{R}}{2}\vol(\Sigma)\,,
\end{equation}
and therefore to the subregion complexity of \cite{Alishahiha:2015rta}. 
We will study this quantity with the normalization \eqref{eq:subregion complexity} of \cite{Abt:2017pmf}.

In \cite{Abt:2017pmf}, we stated that the volume formula \eqref{eq:volume formula} gives an integral expression for $\vol(\Sigma)$ involving only entanglement entropies. 
Since entanglement entropy is a CFT quantity, this integral expression of the
volume can be understood as a CFT formulation of the holographic subregion complexity. 
In the following we expand on the work of \cite{Abt:2017pmf} in greater detail, deriving explicitly the expression for $\vol(\Sigma)$ in terms of entanglement entropies.

%%%%%%%%%%%%%%%%%%%%%%%%%%%%%%%%%%%%%%%%%%%%%%%%%%%%%%%%%%%%%%%%%%%%%%%%%%%%
\subsection{Subregion Complexity in Terms of Entanglement Entropy}
%%%%%%%%%%%%%%%%%%%%%%%%%%%%%%%%%%%%%%%%%%%%%%%%%%%%%%%%%%%%%%%%%%%%%%%%%%%%
In order to express $\vol(\Sigma)$ in terms of entanglement entropy alone, 
we apply the volume formula \eqref{eq:volume formula} to the region $\Sigma$ lying below the geodesic $(\theta_\Sigma,\alpha_\Sigma)$,
\begin{equation}
\label{eq:vol formula for Sigma}
	\frac{\vol(\Sigma)}{4G_N}
	=
	\frac{1}{2\pi}\int_{\cK}\lambda_\Sigma\omega \,.
\end{equation} 
Since we are considering vacuum states, the Crofton form $\omega$ depends only on entanglement entropies (see \eqref{eq: ds and omega for symmetric spaces}). 
The focus of our attention will thus be the chord length $\lambda_\Sigma$. 
For a given geodesic $(\theta,\alpha)$, $\lambda_\Sigma(\theta,\alpha)$ is the length of the segment of $(\theta,\alpha)$ contained in $\Sigma$.
Since $\Sigma$ is convex, this length is simply the geodesic distance between the intersection points $p$, $p'$ of the geodesic $(\theta,$ $\alpha)$ with the boundary of $\Sigma$ (see \figref{fig: different chords}).
We gave in \eqref{eq: geodesic distance} an expression for the geodesic distance between two bulk points in terms of kinematic space quantities,
\begin{equation}
\label{eq: chord length for Sigma}
	\frac{\lambda_\Sigma}{4G_N}
	=
	\frac{1}{4}\int_{\Delta_{pp'}}\omega\,.
\end{equation} 
The set $\Delta_{pp'}(\theta,\alpha)\subset\cK$ is the region bounded by the two point curves corresponding to $p$ and $p'$ for fixed geodesic $(\theta, \alpha)$ (see \figref{fig:point curves}). 
Of course, if $(\theta,\alpha)$ does not intersect $\Sigma$ then $p,p'$ do not exist, and $\Delta_{pp'}$ is empty.
In this case, \eqref{eq: chord length for Sigma} implies $\lambda_\Sigma(\theta,\alpha)=0$ as required. 
Combining \eqref{eq:vol formula for Sigma} and
\eqref{eq: chord length for Sigma}, we obtain an expression for $\vol(\Sigma)$ in terms of entanglement entropy,
\begin{equation}\label{eq: Volume}
	\frac{\vol(\Sigma)}{4G_N^2}
	=
	\frac{1}{2\pi}\int_\cK\omega\Big(
		\int_{\Delta_{pp'}}\omega
		\Big)
	=
	\frac{1}{8\pi}\int_{\cK}d\theta d\alpha
		\int_{\Delta_{pp'}}d\theta'd\alpha'
			 \partial_\alpha^2S(\alpha)\partial_{\alpha'}^2S(\alpha')\,.
\end{equation} 
Applying \eqref{eq:subregion complexity} and inserting the relations $\cR=-2/L^2$ and $G_N=3L/2c$ gives an expression for the subregion complexity in terms of entanglement entropy: 
\begin{equation}
\label{eq: complexity i.t.o. CFT quantities}
	\cC(\theta_\Sigma,\alpha_\Sigma)
	=
	\frac{9}{8\pi c^2}
	\int_{\cK}d\theta\, d\alpha
		\int_{\Delta_{pp'}}d\theta'd\alpha'
			 \partial_\alpha^2S(\alpha)\partial_{\alpha'}^2S(\alpha')\,.
\end{equation}
This expression is one of the main results of this paper: it defines a CFT quantity depending only on $S$ and the integration region $\Delta_{pp'}$. 
To give a purely field theory expression for subregion complexity, it only remains to construct $\Delta_{pp'}$ itself within field theory. 
This will be our next step.

%%%%%%%%%%%%%%%%%%%%%%%%%%%%%%%%%%%%%%%%%%%%%%%%%%%%%%%%%%%%%%%%%%%%%%%
\subsection{Regions of Integration for Complexity}
\label{sec: regions of integration for complexity}
%%%%%%%%%%%%%%%%%%%%%%%%%%%%%%%%%%%%%%%%%%%%%%%%%%%%%%%%%%%%%%%%%%%%%%%
The integrand on the right hand side of \eqref{eq: complexity i.t.o. CFT quantities}  contains only field theory quantities. 
We did, however, use the bulk geometry to construct the region of integration $\Delta_{pp'}(\theta,\alpha)$ for each geodesic $(\theta,\alpha)$. 
Let us now discuss the explicit form of $\Delta_{pp'}$, and show how to construct it directly within CFT. 
Keep in mind that, as discussed in \secref{sec: review of kinematic space},
the geometry of kinematic space can be constructed from entanglement entropy. 
Therefore, if we can construct the $\Delta_{pp'}$ only in terms of the geometry of $\cK$, we no longer reference the bulk explicitly. 
The regions $\Delta_{pp'}$ are bounded by point curves. As
pointed out in \cite{Czech:2014ppa} point curves are space- or light-like geodesics in $\cK$. 
So they are related in a very natural way to the geometry of kinematic
space. 
Thus the only thing left to do is to find a construction rule for the
point curves (i.e. geodesics in $\cK$) of interest that can be formulated from the CFT perspective.

%--------------------------------------------
\begin{figure}[t]
\begin{minipage}{.5\textwidth}
\centering
\includegraphics[scale=0.15]{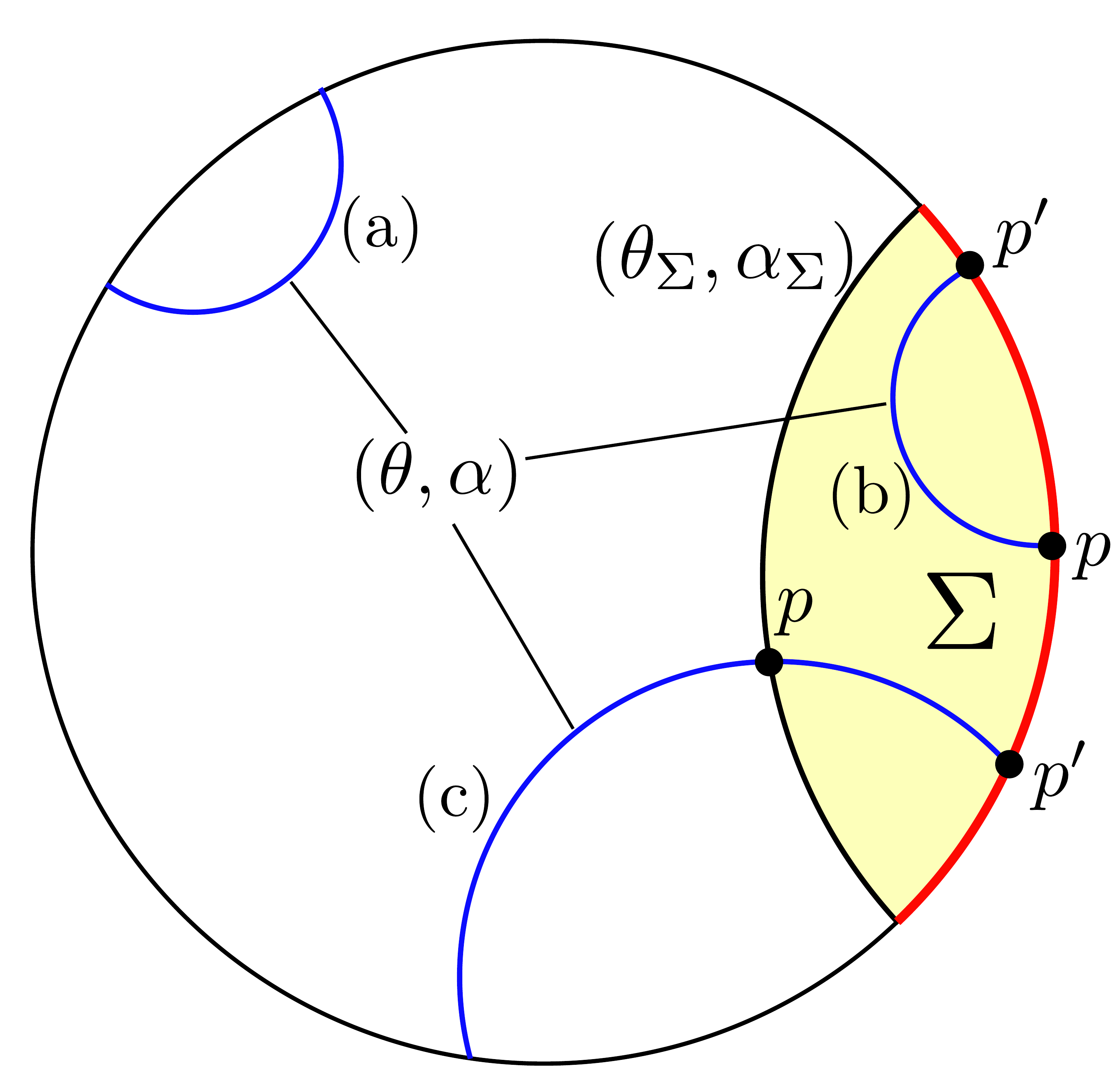}
\end{minipage}%
\begin{minipage}{.5\textwidth}
\includegraphics[scale=0.25]{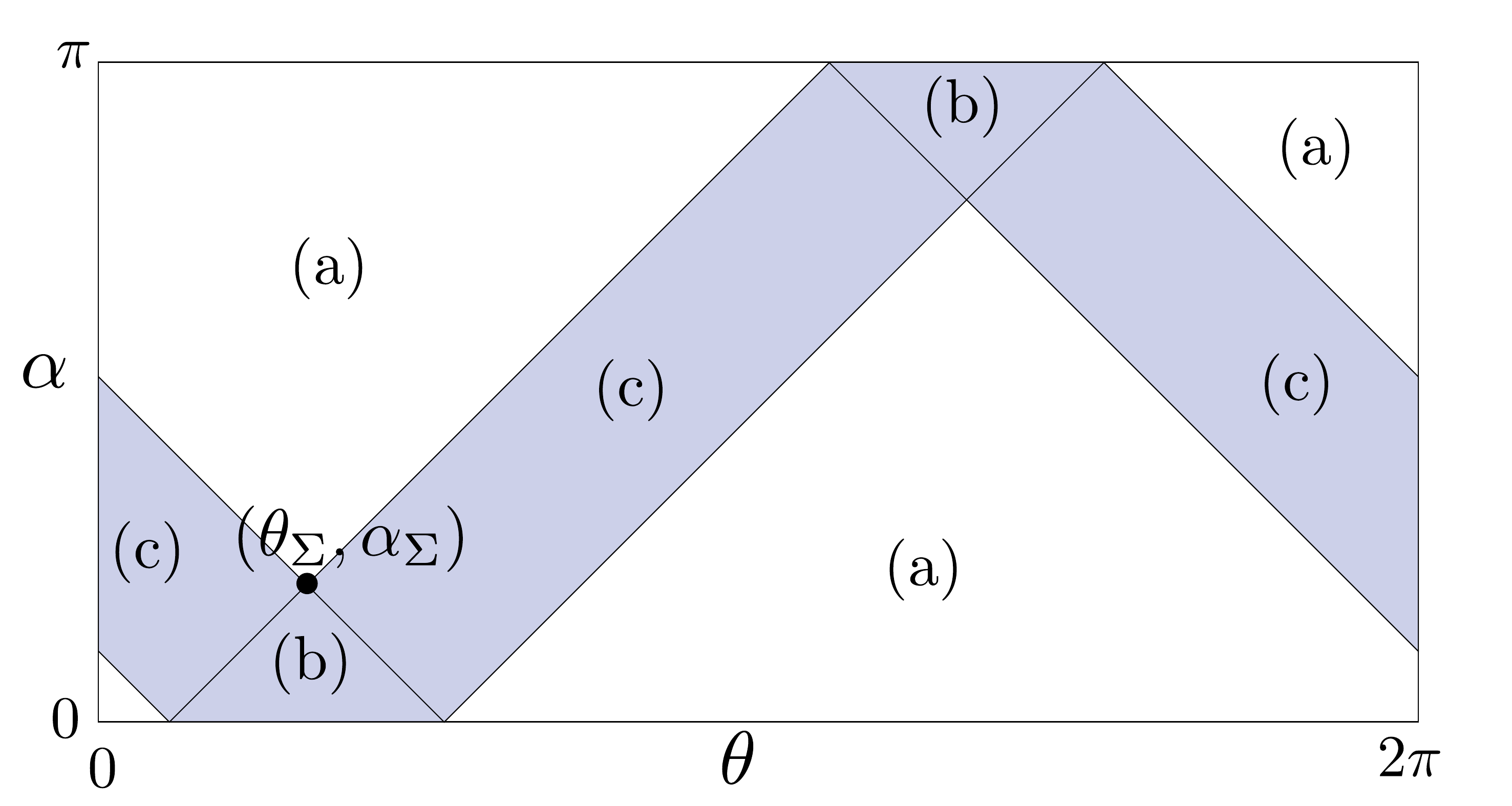}
\end{minipage} 
\caption{In order to construct $\Delta_{pp'}$
			we need to distinguish three different types of geodesics.
			Geodesics of type (a) do not intersect $\Sigma$ at all.
			Type (b) geodesics completely lie inside of $\Sigma$.
			Type (c) geodesics lie only partially inside of $\Sigma$.
			The intersection points $p$ and $p'$ of a geodesic with the 
			boundary can be interpreted as endpoints of entangling regions. 
			On the LHS we show these three types in the bulk, while on the 
			RHS we show where the geodesics of different types are located in 
			kinematic space.}
\label{fig: different chords} 
\end{figure}

%--------------------------------------------

\begin{figure}
\begin{center}
\includegraphics[scale=0.22]{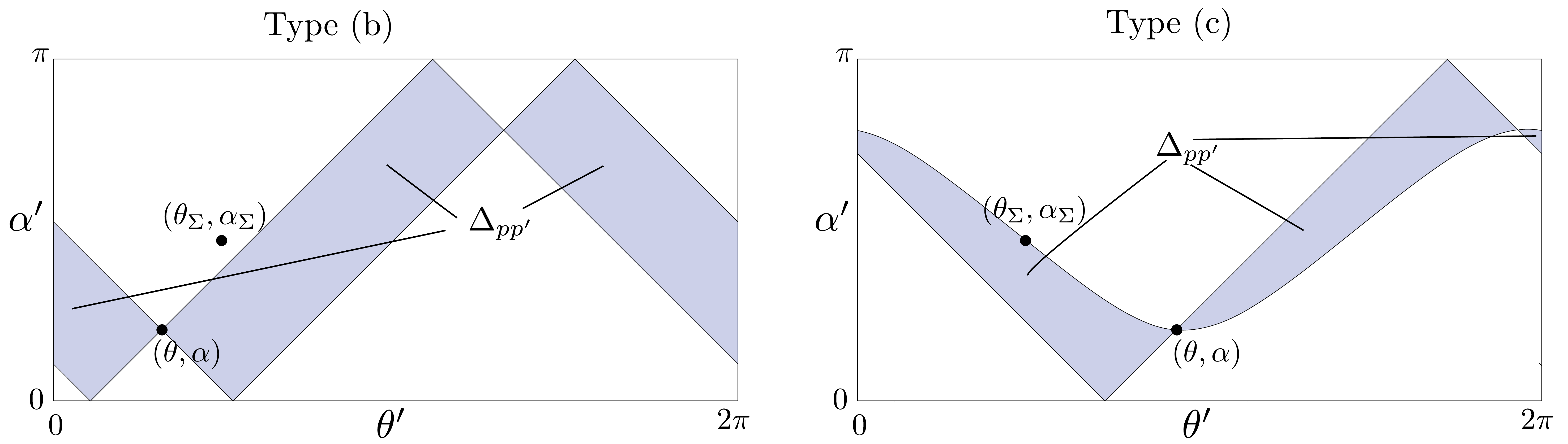} 
\end{center}
\caption{The regions of integration $\Delta_{pp'}(\theta,\alpha)$ for geodesics
			$(\theta,\alpha)$ that are of type (b) and (c) w.r.t.
			$(\theta_\Sigma,\alpha_\Sigma)$. For type (b) geodesics $\Delta_{pp'}$ is
			bounded by light rays. For type (c) geodesics one boundary of
			$\Delta_{pp'}$ is the unique point curve that passes through
			$(\theta_\Sigma,\alpha_\Sigma)$ and $(\theta,\alpha)$.}
\label{fig: type b c}
\end{figure}

We first examine these point curves from the bulk point of view and then translate our results into CFT language. 
We distinguish three types of geodesics, as depicted in \figref{fig: different chords}:
\begin{description}
\item[Type (a)] 
geodesics are those $(\theta,\alpha)$ that do not intersect $\Sigma$ at all. 
Such geodesics have $\Delta_{pp'}=\emptyset$, and therefore $\lambda_\Sigma(\theta,\alpha)=0$.

\item[Type (b)]
geodesics are those $(\theta,\alpha)$ that lie completely inside of $\Sigma$.
In this situation, the intersection points $p$ and $p'$ are located on the conformal boundary, i.e. the constant time slice of the CFT. 
They are the endpoints of the entangling interval associated with $(\theta,\alpha)$ and can be interpreted as points that lie on the boundary of $\cK$. In particular they lie within the entangling
interval corresponding to $(\theta_\Sigma,\alpha_\Sigma)$. 
In this case the corresponding point curves are null geodesics \cite{Czech} emitted from $p$ and $p'$.
Consequently, the region $\Delta_{pp'}$ enclosed by these light rays consists of causal diamonds in $\cK$. 
An example of such a $\Delta_{pp'}$ is depicted in \figref{fig: type b c}.

\item[Type (c)]
geodesics are those $(\theta,\alpha)$ that lie only partially inside $\Sigma$. 
In this case, one of the intersection points $p$ lies on the geodesic
$(\theta_\Sigma,\alpha_\Sigma)$, while the other, $p'$, lies on the boundary in the interval specified by $(\theta_\Sigma,\alpha_\Sigma)$. 
As for type (b), $p'$ is one of the endpoints of the entangling region corresponding to $(\theta,\alpha)$.
Therefore, treating $p'$ as a boundary point of $\cK$, the point curve of $p'$ is once again a null geodesic emitted from $p'$. 
As mentioned in section \ref{sec: review of kinematic space}, the point curve of $p$ is a space-like geodesic in $\cK$. 
Noting that $p$ lies on both geodesics $(\theta_\Sigma,\alpha_\Sigma)$ and $(\theta,\alpha)$, the point curve of $p$ is determined to be the unique geodesic in $\cK$ containing both $(\theta_\Sigma,\alpha_\Sigma)$ and $(\theta,\alpha)$. 
One such $\Delta_{pp'}$ is depicted in \figref{fig: type b c}.
\end{description}

%--------------------------------------------
%--------------------------------------------

Note that it is not possible for both $p$ and $p'$ to lie on the geodesic $(\theta_\Sigma,\alpha_\Sigma)$, since this would mean that the geodesic $(\theta,\alpha)$ intersects $(\theta_\Sigma,\alpha_\Sigma)$ twice.
Therefore, types (a)-(c) exhaust all possibilities. 
\figref{fig: different chords} illustrates the location of different types of geodesics in kinematic space. 
Type (b) geodesics lie in the past of $(\theta_\Sigma,\alpha_\Sigma)$ and the future of $(\theta_\Sigma+\pi,\pi-\alpha_\Sigma)$, while %the set of 
type (c) geodesics 
%is 
are those
enclosed by the light rays emitted from the endpoints of the entangling region associated to $(\theta_\Sigma,\alpha_\Sigma)$. 
All other geodesics are of type (a).

We constructed the region of integration $\Delta_{pp'}$ by identifying the point curves of $p$ and $p'$.
The next step is to reformulate this construction in terms of CFT objects.
Recall that $\cK$ has an interpretation as the space of CFT intervals.
Through this $\vol(\Sigma)$ acquires meaning without reference to the bulk geometry.
When we consider $(\theta,\alpha)$ and $(\theta_\Sigma,\alpha_\Sigma)$ to be entangling intervals, the three types (a)-(c) distinguish where the endpoints of $(\theta,\alpha)$ lie relative to $(\theta_\Sigma,\alpha_\Sigma)$ (see \figref{fig: different chords}):
an entangling interval is of type (a) if none of its endpoints lie inside $(\theta_\Sigma,\alpha_\Sigma)$; 
the intervals with both endpoints lying inside $(\theta_\Sigma,\alpha_\Sigma)$ are of type (b);
of type (c) are the entangling regions with only one endpoint lying in $(\theta_\Sigma,\alpha_\Sigma)$.

We have therefore constructed $\Delta_{pp'}$ using only entangling regions and the geometry of $\cK$:
\begin{itemize}
	\item If $(\theta,\alpha)$ is of type (a), we set 
			$\Delta_{pp'}(\theta,\alpha)=\emptyset$.
	\item If $(\theta,\alpha)$ is of type (b),
			$\Delta_{pp'}(\theta,\alpha)$
			is the region bounded by the light rays emitted from both boundary
			points of $(\theta,\alpha)$. (\figref{fig: type b c})
	\item If $(\theta,\alpha)$ is of type (c),
			$\Delta_{pp'}(\theta,\alpha)$
			is the region bounded by the light rays emitted from the endpoint of
			$(\theta,\alpha)$ that lies inside of $(\theta_\Sigma,\alpha_\Sigma)$ and
			the space-like geodesic that intersects $(\theta_\Sigma,\alpha_\Sigma)$
			and $(\theta,\alpha)$. (\figref{fig: type b c})
\end{itemize}  

We now have a formula specified by two components: the geometry of kinematic space, and the integration regions $\Delta_{pp'}$. 
The geometry of $\cK$ is defined in terms of entanglement entropy, while we have shown that the form $\Delta_{pp'}$ is determined by this geometry.
The resulting object \eqref{eq: complexity i.t.o. CFT quantities} is therefore defined for any CFT, regardless of whether it has a holographic dual or not.
Our construction shows, however, that when the CFT does possess a weakly curved holographic dual, this quantity coincides with the holographic subregion complexity \eqref{eq:subregion complexity}.

We emphasize that the only entangling intervals contributing to \eqref{eq: complexity i.t.o. CFT quantities} are those with one or both endpoints lying in the interval $(\theta_\Sigma,\alpha_\Sigma)$.
In other words, only intervals of type (b) and (c) are present.
For the outer integral (over $\theta,\alpha$) this is clear, since $\Delta_{pp'}(\theta,\alpha)$ is empty for intervals with no endpoint contained in $(\theta_\Sigma,\alpha_\Sigma)$. 
To see this for the integral computing chord lengths (over $\theta',\alpha'$), note that the region of integration $\Delta_{pp'}(\theta,\alpha)$ for type (b) and (c) is given by the set of geodesics passing through the chord of geodesic $(\theta,\alpha)$ in $\Sigma$ (see \secref{sec: review of kinematic space}).
As a result, the geodesics in $\Delta_{pp'}(\theta,\alpha)$ intersect $\Sigma$ and are thus of type (b) or (c) as well.

\medskip
Let us briefly consider the more general problem of evaluating the volume of an arbitrary bulk region $Q$.
It can be expressed in terms of entanglement entropies using the same basic procedure as above:
one merely applies the formula for geodesic distances \eqref{eq: geodesic distance} to the chord length $\lambda_Q$. 
The drawback is that the bulk region is no longer bounded by geodesics, making the regions of integration in kinematic space difficult to determine without explicit knowledge of the bulk. 
We still wish to stress, however, that it is possible to express arbitrary volumes in terms of entanglement entropies, in the same way that it is possible to express the length of an arbitrary curve as an integral over kinematic space.

%%%%%%%%%%%%%%%%%%%%%%%%%%%%%%%%%%%%%%%%%%%%%%%%%%%%%%%%%%%%%%%%%%%%%%%%%%%%%%%%%
\subsection{Subregion Complexity for Global AdS$_3$}
\label{sec: subregion complexity}
%%%%%%%%%%%%%%%%%%%%%%%%%%%%%%%%%%%%%%%%%%%%%%%%%%%%%%%%%%%%%%%%%%%%%%%%%%%%%%%%%
The last section explained how to construct the regions of integration in \eqref{eq: complexity i.t.o. CFT quantities} from the field theory perspective. 
We now evaluate \eqref{eq: complexity i.t.o. CFT quantities} to compute subregion complexities. 
In this section we consider global AdS$_3$ \eqref{eq: metric global AdS3} and present the complexity for the cases where (1) the entangling interval $(\theta_\Sigma,\alpha_\Sigma)$ is the entire CFT circle, and (2) where it is half of this circle.
General intervals for the Poincar\'e patch will be considered in the next section.

\begin{figure}
\begin{minipage}{.5\textwidth}
\centering
\includegraphics[scale=0.22]{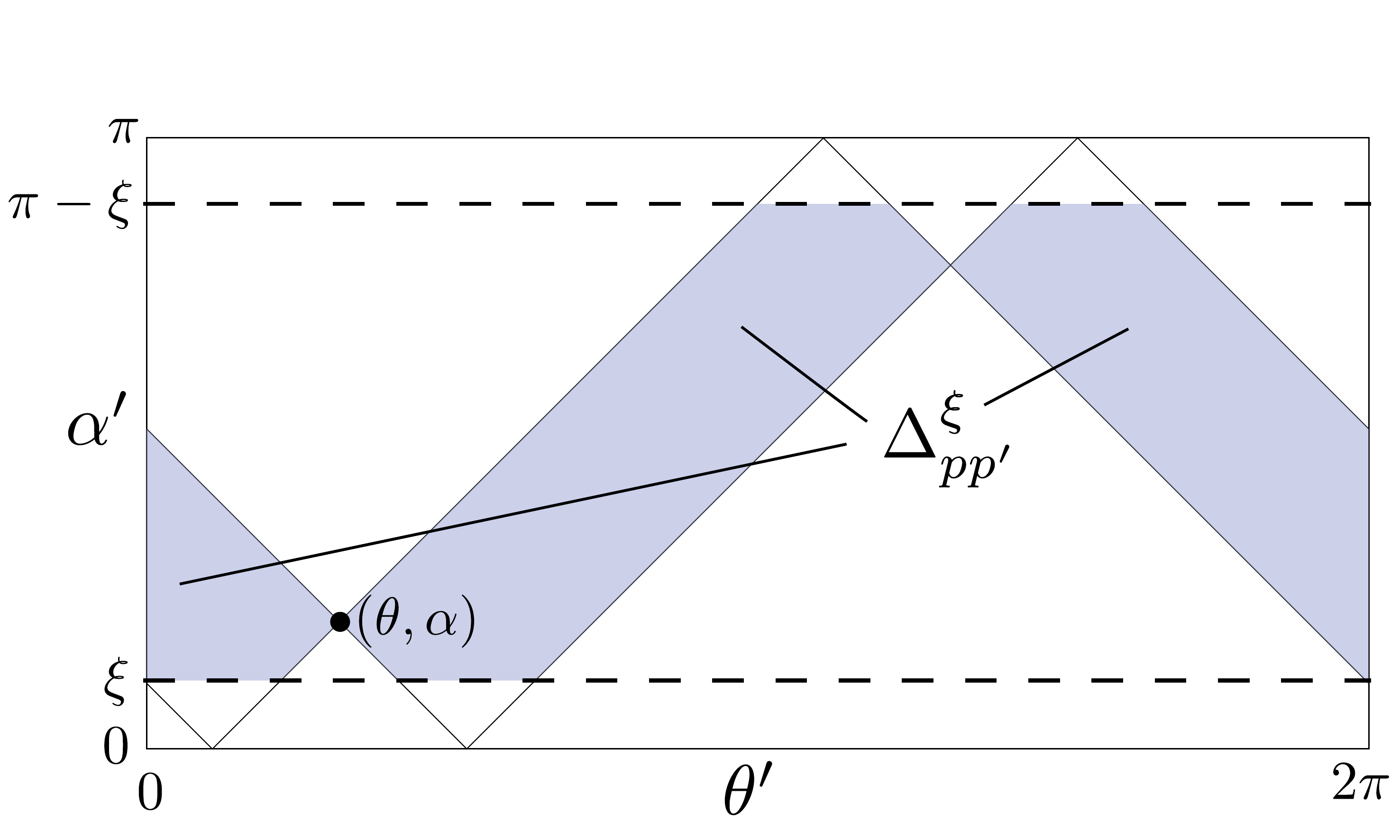} 
\end{minipage}%
\begin{minipage}{.5\textwidth}
\centering
\begin{tikzpicture}[scale=0.8]
  \draw (-0.1231,1.5643) arc (9:15:10cm);
  \draw[red] (-0.1231,1.5643) arc (9:-15:10cm);
  \draw[dashed] (-1,0) arc (0:16:9cm);
  \draw[dashed] (-1,0) arc (0:-15.5:9cm);
  \draw[blue] (-0.0381,0.8716) arc (240:210:4cm);
  \fill[\figyellow] (-3,-2) arc (160:131.5:7cm) arc (5.5:-15:9cm);
  \draw (-3,-2) arc (160:122:7cm);
  \draw (-2,-1) node {$\Sigma_\epsilon$};
  \draw[->] (-3.5,0) -- (-1,0);
  \draw (-2.5,0) node[above] {$r_\epsilon$};
\end{tikzpicture}
\end{minipage}
\caption{LHS: The region of integration for type (b) intervals.
  By construction the region of integration
  $\Delta_{pp'}(\theta,\alpha)$ of a type (b) interval
  $(\theta,\alpha)$ consists of causal diamonds.
  After introducing cutoffs at $\alpha'=\xi$ and $\alpha'=\pi-\xi$
  the region of integration reduces to
  $\Delta_{pp'}^\xi(\theta,\alpha)$.
  RHS: Close-up of the region near the edge of $\Sigma$ showing the
  inequivalence of radial and kinematic space cutoff schemes.
  By choosing a cutoff at a fixed radial coordinate $r_\epsilon$ in
  the bulk (dashed line) we reduce $\Sigma$ to a regularized region
  $\Sigma_\epsilon$ (yellow) whose volume is to be computed.
  The blue geodesic does not contribute to $\vol(\Sigma_\epsilon)$,
  but it contributes to the volume regularized with the kinematic
  space cutoff scheme, since its size is larger than $\xi$.}
\label{fig: type b chords}
\end{figure}
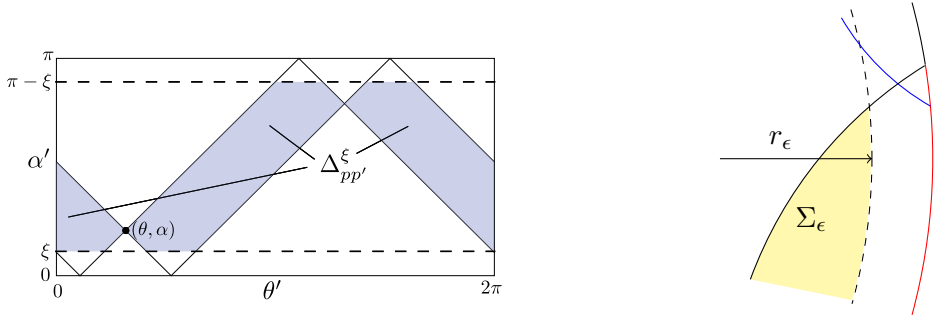

Consider equation \eqref{eq: complexity i.t.o. CFT quantities} for the subregion complexity. 
The entanglement entropy $S$ is given by \eqref{eq: EE for global AdS3}.
Note that $S\propto c$ and thus $\cC\propto c^0$.
The complexity diverges, and must be regularized. 
In the bulk, subregion complexity is identified with the volume below the RT surface. 
Usually a radial cutoff is chosen to compute this volume.
We could translate this cutoff to kinematic space and use it for our computations. 
However, this regularization is not very natural from the kinematic space or CFT perspectives.
Once more we emphasize that we wish to compute complexity without using the bulk. 
We therefore choose a different cutoff scheme: We introduce a minimal opening angle $\xi$ and only work with the part of kinematic space with opening angles $\alpha,\alpha'\in[\xi,\pi-\xi]$ as depicted in the LHS of \figref{fig: type b chords}.
From the CFT perspective this means that we are only working with entangling intervals with an opening angle larger than $\xi$, and whose complement has opening angle larger than $\xi$.

If we take the entangling region to be the entire constant time slice, all entangling intervals $(\theta,\alpha)$ are of type (b) (see \secref{sec: regions of integration for complexity}), and therefore $\Delta_{pp'}(\theta,\alpha)$ consists of causal diamonds that now need to be cut off at $\alpha'=\xi$ and $\alpha'=\pi-\xi$.
The resulting complexity of the entire CFT circle is thus
\begin{equation}
\label{eq: complexity circle}
	\cC(\text{circle})
	=
	\frac{9}{8\pi c^2}\int_0^{2\pi}d\theta\int_\xi^{\pi-\xi} d\alpha
		\int_{\Delta^\xi_{pp'}}d\theta'd\alpha'
			 \partial_\alpha^2S(\alpha)\partial_{\alpha'}^2S(\alpha')\,.
\end{equation}
The region of integration $\Delta^\xi_{pp'}$ is depicted in the LHS of \figref{fig: type b chords}.
%--------------------------------------------
%--------------------------------------------
It is easy to verify that
\begin{equation}
\label{eq: ent entropy from kin sp}
	\int_{\Delta^\xi_{pp'}}d\theta'd\alpha'
			 \partial_{\alpha'}^2S(\alpha')
	=
	-\frac{8c}{3}\left(
		\log\left(\frac{\sin(\alpha)}{\sin(\xi)}\right)
		+
		\xi\cot(\xi)\right)
	=
	-\frac{8c}{3}\left(\log\left(
		\frac{\sin(\alpha)}{\xi}\right)
		+
		\cO(\xi^0)\right) \,.
\end{equation}
As $\xi\rightarrow 0$, this integral approaches $-8$ times the entanglement entropy of the boundary interval $[p,p']$, provided the CFT cutoff is identified with $\xi$ appropriately.
The integral \eqref{eq: ent entropy from kin sp} gives, up to a constant prefactor, the length of the geodesic connecting $p$ and $p'$ (see \eqref{eq: geodesic distance} and \figref{fig: different chords}).
Via the RT proposal, this length is associated with entanglement entropy.
So we see that in \eqref{eq: ent entropy from kin sp} we indeed obtain the correct logarithmic divergence in our chosen cutoff scheme.

By inserting \eqref{eq: ent entropy from kin sp} into
\eqref{eq: complexity circle} we obtain
\begin{equation}
\label{eq: complexity circle result}
	\cC(\text{circle})
	=
	4\left(\xi\cot^2(\xi)+\cot(\xi)+\xi-\frac{\pi}{2}\right)
	=
	\frac{8}{\xi}-2\pi+\cO(\xi^2)\,.
\end{equation}
In \cite{Abt:2017pmf} the complexity was determined by computing the volume below the RT surface directly. This
computation used the radial cutoff $r_{\epsilon}=L\lcft/\epsilon$,  where $\lcft$ is the radius of the CFT circle and $\epsilon$ is the UV cutoff.
We can match our result for the divergent and constant parts of the complexity with those presented in
\cite{Abt:2017pmf} by setting $\xi=4\epsilon/\pi \lcft$.  
Just as in \cite{Abt:2017pmf} we obtain the constant part $-2\pi$.

%Note that this is accomplished by using a cutoff scheme in kinematic space that cannot be associated with a geometric cutoff in the bulk.
We emphasize that the kinematic space cutoff scheme we have used is not equivalent to any sharp geometric cutoff in the bulk.
%To be more precise: 
%Choosing a cutoff $\xi$ in $\cK$ does not correspond to choosing some cutoff curve in the bulk. 
%In particular, it is not equivalent to a radial cutoff. 
To see this explicitly, we consider the region $\Sigma_\epsilon$ obtained by regulating $\Sigma$ at the radial cutoff $r_\epsilon$, as shown in the RHS of \figref{fig: type b chords}.
% corresponds to computing the volume of the regularized bulk region $\Sigma_\epsilon$
When computing the regularized subregion complexity in the kinematic space prescription with a cutoff at fixed $\xi$, however, the result receives contributions from geodesics---like the blue geodesic of the figure---that have an opening angle larger than $\xi$ and yet do not intersect the bulk region $\Sigma_\epsilon$.
%As a result, the kinematic space regularization prescription used above generally yields a value distinct from $\vol(\Sigma_\epsilon)$.
%the corresponding kinematic space cutoff $\xi=4\epsilon/\pi\ell_{CFT}$ does not lead to the computation of $\vol(\Sigma_\epsilon)$ 
%due to the contributions of geodesics with opening angle larger than $\xi$ that do not intersect the bulk region $\Sigma_\epsilon$ (exemplified by the blue geodesic).
%It is not possible to translate the cutoff scheme in $\cK$ into a bulk cutoff scheme in a simple geometric fashion.

The fact that the constant coefficient in the subregion complexity is the same in both cutoff schemes supports the idea that it is indeed universal \citep{Alishahiha:2015rta, Abt:2017pmf}. 
This statement is corroborated by the result for the complexity of one half of the CFT circle, computed in Appendix \ref{ap: subregion complexity for the semicircle}.
We find
\begin{equation}
\label{eq: complexity for semicircle}
	\cC(\text{semicircle})
	=
	2\xi\cot^2(\xi)+2\cot(\xi)+2\xi-\pi
	=
	\frac{4}{\xi}-\pi+\cO(\xi^2)\,.
\end{equation}
The constant and divergent parts of the complexity match the results of \cite{Abt:2017pmf} provided we identify $\xi=4\epsilon/\pi \lcft$.

%%%%%%%%%%%%%%%%%%%%%%%%%%%%%%%%%%%%%%%%%%%%%%%%%%%%%%%%%%%%%%%%%%%%%%%%%%%%%%%%
\subsection{Subregion Complexity for the Poincar\'e Patch}
\begin{figure}
  \begin{tabular}{ccc}
    \begin{tikzpicture}
      \draw[fill=\figblue] (1.5,0) arc(0:180:1.5cm) -- (-1.5,0);
      \draw[->] (-2,0) -- (2,0) node[right] {$x$};
      \draw[->] (-1.9,0) -- (-1.9,2.5) node[right] {$z$};
      \draw (-1,0) arc(0:35:3cm) node[above,right] {$\lambda_{\Sigma,(c),+}$};
      \draw (0.5,0) arc(0:180:0.5cm) -- (-0.5,0);
      \draw (0,0.5) node[above] {$\lambda_{\Sigma,(b)}$};
      \draw (1.4,1.29) node[above] {$\lambda_{\Sigma,(c),-}$} arc(140:180:2cm);
    \end{tikzpicture} &
    \begin{tikzpicture}[scale=0.75]
      \fill[color=\figblue] plot[domain=-2.75:-0.583,smooth,variable=\x] ({\x},{sqrt(1+(\x+1)*(\x+1))}) -- (-2.75,3.25) -- (-3,2.236);
      \fill[color=\figblue] plot[domain=-0.583:2.25,smooth,variable=\x] ({\x},{sqrt(1+(\x+1)*(\x+1))}) -- (3,2.5) -- (1,0.5) -- (0,0.5) -- (-0.583,1.083);      \draw[->] (-3,0) -- (3,0) node[right] {$\tilde \chi$};
      \draw[->] (0,0) -- (0,3) node[above] {$\tilde\psi$};
      \draw plot[domain=-3:2.25,smooth,variable=\x] ({\x},{sqrt(1+(\x+1)*(\x+1))});
      \draw (2.25,3.4) node[right] {$\tilde\psi_A$};
      \draw[domain=0.5:3,smooth,variable=\x] plot ({\x},{\x-0.5});
      \draw (3,2.5) node[right] {$\tilde\psi_B$};
      \draw[domain=-2.75:0.5,smooth,variable=\x] plot ({\x},{-\x+0.5});
      \draw[dashed] (-3,0.5) -- (3,0.5) node[right] {$\xi$};
    \end{tikzpicture} &
    \begin{tikzpicture}[scale=0.75]
      \fill[color=\figblue] (-0.5,3) -- (1.5,1) -- (1,0.5) -- (0,0.5) -- (-1.5,2);
      \fill[color=\figblue] (3.5,3) -- (1.5,1) -- (2,0.5) -- (3,0.5) -- (4.5,2);
      \fill[color=\figblue] (1.5,1) -- (2,0.5) -- (1,0.5) -- (1.5,1);
      \draw[dashed] (-1.5,0.5) -- (4.5,0.5) node[right] {$\xi$};
      \draw (2.5,0) -- (-0.5,3);
      \draw (0.5,0) -- (3.5,3);
      \draw (2.5,0) -- (4.5,2);
      \draw (0.5,0) -- (-1.5,2);
      \draw[->] (-0.7,0) -- (-0.7,3) node[above] {$\psi$};
      \draw[->] (-1.5,0) -- (4.5,0) node[right] {$\chi$};
      \draw (0.5,0) node[below] {$u_\Sigma$};
      \draw (2.5,0) node[below] {$v_\Sigma$};
      \draw (0.4,1.2) node {\footnotesize{$(c),+$}};
      \draw (2.6,1.2) node {\footnotesize{$(c),-$}};
      \draw (1.5,0.7) node {\footnotesize{$(b)$}};
    \end{tikzpicture}    
  \end{tabular}
  \caption{Chord lengths for a subregion $\Sigma$ (left), regions of integration for chord length $\lambda_{\Sigma,(c),+}$ (middle) and for the full volume (right) in kinematic space. In the right-most figure, the regions of integration for the chord lengths are marked with their respective subscripts. Note that as in \secref{sec: volume formula proof} we restrict to positively oriented geodesics ($\psi>0$).}
  \label{fig:chord-length-subregion-poincare}
\end{figure}
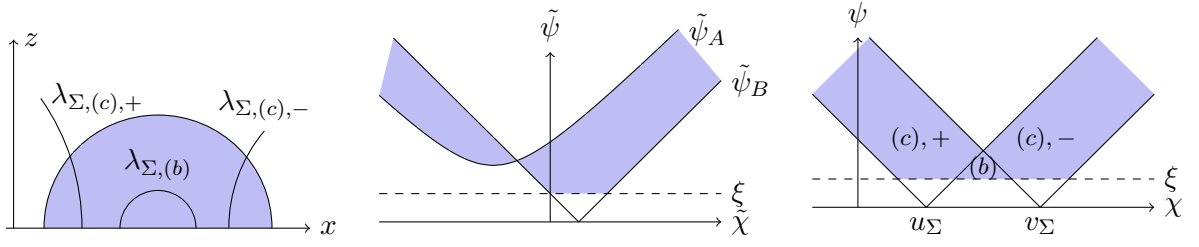

%%%%%%%%%%%%%%%%%%%%%%%%%%%%%%%%%%%%%%%%%%%%%%%%%%%%%%%%%%%%%%%%%%%%%%%%%%%%%%%%
%
%As a further application of \eqref{eq: complexity i.t.o. CFT quantities} 
We now compute the subregion complexity for the Poincar\'e patch using kinematic space. We use the coordinates $(\chi,\psi)$ introduced in \eqref{eq: kin sp coordinates for Pp} for elements in kinematic space, i.e.~entangling intervals. 
To compute the subregion complexity for an interval $(\chi_\Sigma,\psi_\Sigma)$
the corresponding chord lengths have to be calculated.
Recall from \secref{sec: regions of integration for complexity} that they are given by an integral over the area between two point curves $(\tilde\chi,\tilde\psi_A(\tilde\chi))$ and $(\tilde\chi,\tilde\psi_B(\tilde\chi))$.
There are contributions from two types of intervals, type (b) and type (c).
For a type (b) interval $(\chi,\psi)$, both point curves are light rays, 
$\tilde\psi_{A,B}=|\chi\pm\psi-\tilde \chi|$, whereas if $(\chi,\psi)$ is of type (c) one point curve is given by $\tilde\psi_{A,B}=\sqrt{\psi^2+(\tilde \chi-x_\lambda)^2-(x_\lambda-\chi)^2}$. 
Here, 
\be
x_\lambda=\frac{\psi_\Sigma^2-\psi^2+\chi^2-\chi_\Sigma^2}{2(\chi-\chi_\Sigma)}
\ee
is the $x$ coordinate of the intersection point of the geodesics $(\chi_\Sigma,\psi_\Sigma)$ and $(\chi,\psi)$. 
Integrating \eqref{eq: chord length for Sigma} with kinematic space cutoff $\psi=\xi$ yields
\begin{align}
\lambda_{\Sigma,(b)} &= 2L\biggl[\log\Bigl(\frac{\psi}{\xi}\Bigr)+1\biggr] 
\,,\\
\lambda_{\Sigma,(c),\pm} &= \frac{1}{2}\lambda_{\Sigma,(b)}+\frac{L}{2}\log\left(\frac{\psi\pm(\chi-x_\lambda)}{\psi\mp(\chi-x_\lambda)}\right)
\,.
\end{align}
Here, the length $\lambda_{\Sigma,(c),\pm}$ corresponds to a geodesic with only its right (left) endpoint inside the boundary interval (\figref{fig:chord-length-subregion-poincare}).
We have implicitly assumed that for type (c) intervals, the non-lightlike point curve stays above the cutoff for all $\tilde\chi$. 
The error due to this assumption is of order $\cO(\xi)$ and can be ignored. 
\label{sec:pp volume}
Applying \eqref{eq:vol formula for Sigma} the volume is now given by
\begin{multline}
\vol(\Sigma)=\frac{4cG_N}{6\pi} \biggl[
  \int\limits_\xi^{\psi_\Sigma}d\psi\int\limits_{u_\Sigma+\psi}^{v_\Sigma-\psi}\frac{d\chi}{\psi^2}\lambda_{\Sigma,(b)}
  +\int\limits_\xi^{\psi_\Sigma}d\psi\int\limits_{v_\Sigma-\psi}^{v_\Sigma+\psi}\frac{d\chi}{\psi^2}\lambda_{\Sigma,(c),-} \\
  +\int\limits_{\psi_\Sigma}^{\infty}d\psi\int\limits_{u_\Sigma+\psi}^{v_\Sigma+\psi}\frac{d\chi}{\psi^2}\lambda_{\Sigma,(c),-} 
  +\int\limits_\xi^{\psi_\Sigma}d\psi\int\limits_{u_\Sigma-\psi}^{u_\Sigma+\psi}\frac{d\chi}{\psi^2}\lambda_{\Sigma,(c),+}
  +\int\limits_{\psi_\Sigma}^{\infty}d\psi\int\limits_{u_\Sigma-\psi}^{v_\Sigma-\psi}\frac{d\chi}{\psi^2}\lambda_{\Sigma,(c),+} \biggr]\,,
\label{eq:pp volume integral}
\end{multline}
where $u_\Sigma=\chi_\Sigma-\psi_\Sigma$ and $v_\Sigma=\chi_\Sigma+\psi_\Sigma$ are the endpoints of the entangling interval.
Separating $\lambda_{\Sigma,(c),\pm}$ into a divergent part proportional to $\lambda_{\Sigma,(b)}$ and a finite part, we write the divergent part of the volume as 
\begin{equation}
\int_\xi^\infty d\psi\, 2L^2(v_\Sigma-u_\Sigma)\frac{\log(\psi/\xi)+1}{\pi\psi^2}=8L^2\frac{\psi_\Sigma}{\pi\xi}+\mathcal{O}(\xi) \,,
\end{equation}
reproducing the expected divergent behavior. 
For the finite part, we set $\xi$ to zero and evaluate the remaining integrals in \eqref{eq:pp volume integral} to obtain 
\begin{equation}
\int_0^\infty d\psi\, \frac{2L^2}{\pi\psi^2}\left[\psi_\Sigma\log\left|\frac{\psi_\Sigma^2}{\psi_\Sigma^2-\psi^2}\right|+\psi\log\left|\frac{\psi_\Sigma-\psi}{\psi_\Sigma+\psi}\right|\right]+\mathcal{O}(\xi)=-L^2\pi+\mathcal{O}(\xi) \,.
\end{equation}
By applying this result to \eqref{eq:subregion complexity} and using $\mathcal{R}=-\frac{2}{L^2}$ and $G_N=\frac{3L}{2c}$ we find the subregion complexity to be
\begin{equation}
  \cC(\chi_\Sigma,\psi_\Sigma)=\frac{8\psi_\Sigma}{\pi\xi}-\pi+\cO(\xi) \,.
\end{equation}
As expected, up to subleading differences due to the choice of cutoff scheme we reproduce the complexity as computed by direct bulk integration using a cutoff $\epsilon$ in the $z$ coordinate, $\cC(\chi_\Sigma,\psi_\Sigma) = \frac{2\psi_\Sigma}{\epsilon}-\pi+\cO(\epsilon)$ \cite{Abt:2017pmf}. 
Interestingly, the finite part comes only from type (c) geodesics, whereas the divergent part requires contributions from both type (b) and type (c) geodesics 
to get the correct result.

It is also possible to implement a sharp bulk cutoff at $z=\epsilon$ in the kinematic space formalism, which is equivalent to using only point curves lying completely above $\psi=\epsilon$. 
In this case, the agreement is exact to all orders in $\epsilon$.

%%%%%%%%%%%%%%%%%%%%%%%%%%%%%%%%%%%%%%%%%%%%%%%%%%%%%%%%%%%%%%%%%%%%%%%%%%%%%%%%
\subsection{Mutual Information and the Volume Formula in the Poincar\'e Patch}
\label{sec: chord lenghts for Pp and mi}
%%%%%%%%%%%%%%%%%%%%%%%%%%%%%%%%%%%%%%%%%%%%%%%%%%%%%%%%%%%%%%%%%%%%%%%%%%%%%%%%
We conclude our analysis of the Poincar\'e patch with a reformulation of \eqref{eq: complexity i.t.o. CFT quantities}.
The paper \cite{Wen:2018whg} showed that the length of a sufficiently long geodesic chord can be interpreted in terms of a mutual information in the dual CFT.
We now apply this observation to the chord lengths $\lambda_\Sigma$ appearing in the volume formula \eqref{eq:vol formula for Sigma}. 

\begin{figure}
\begin{center}
\includegraphics[scale=0.3]{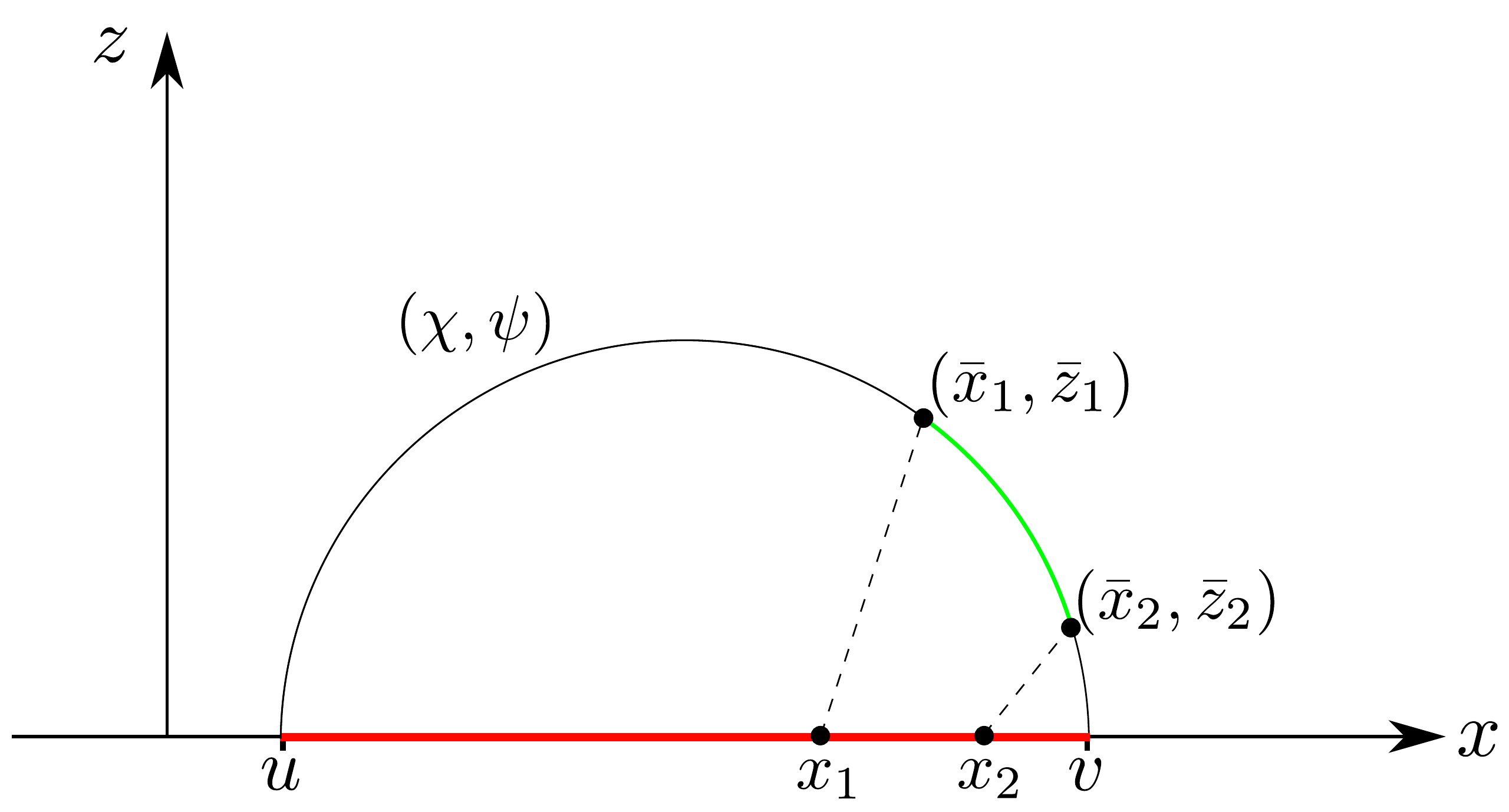} 
\end{center}
\caption{Points on the geodesic $(\chi,\psi)$ can be assigned to points on the corresponding entangling interval, depicted in red. This allows to write the length of the segment of the geodesic (green) lying between the two points $(\bar{x}_1,\bar{z}_1)$ and $(\bar{x}_2,\bar{z}_2)$ in terms of entanglement entropies.}
\label{fig: points associated with each other}
\end{figure}

Consider a Poincar\'e patch geodesic $(u,v)=(\chi-\psi,\chi+\psi)$ in the notation of \secref{sec:pp volume}. 
Based on the bulk modular flow (equivalent to time evolution in the hyperbolic slicing of \cite{Myers:2010xs,Myers:2010tj}), \cite{Wen:2018whg} assigned to each point $(\bar{x},\bar{z})$ on the geodesic a point $x$ in the corresponding entangling interval (\figref{fig: points associated with each other}),%
\footnote{I.e. $(\bar{x},\bar{z})$ satisfies $\bar{z}^2=\psi^2-(\bar{x}-\chi)^2$.}
\begin{equation}
\label{eq: asso of bulk points with bdy points}
	x
	=
	\frac{\psi^2-\sqrt{\psi^4-(\bar{x}-\chi)^2\psi^2}}{\bar{x}-\chi}+\chi\,.
\end{equation}
%
%--------------------------------------------
%--------------------------------------------
%
The length of the geodesic segment between the two points $(\bar{x}_1,\bar{z}_1)$ and $(\bar{x}_2,\bar{z}_2)$ is given by 
\begin{equation}
\label{eq: geodesic distance poincare}
	d\bigl((\bar{x}_1,\bar{z}_1), (\bar{x}_2,\bar{z}_2)\bigr)
	=
	L\log\Big(\frac{2\psi(x_2-x_1)}{(x_1-u)(v-x_2)}+1\Big)
	=
	L\log\eta \,,
\end{equation}
where $x_i$ ($i=1,2$) is the boundary point assigned to $(\bar{x}_i,\bar{z}_i)$, and we take $x_2>x_1$ by convention. 
The last expression is in terms of the conformal cross-ratio
\be
\eta = \frac{(v-x_1)(x_2-u)}{(v-x_2)(x_1-u)} \,.
\ee
The setup is depicted in \figref{fig: points associated with each other}. 
Expressing the chord length $\lambda_\Sigma$ in the form \eqref{eq: geodesic distance poincare}, the volume formula \eqref{eq:vol formula for Sigma} takes a compact form in terms of the cross ratio: 
\begin{equation}
\label{eq: C i.t.o eta}
	\mathcal{C}(\chi_\Sigma,\psi_\Sigma)
	= 
	\frac{3}{\pi c}
	\int_{\text{types}\,b,c}\log \eta\,\omega\,.
\end{equation}
For type (b) geodesics, the chord length reduces to the RT formula.
For type (c) geodesics, one of the points $(\bar{x}_i,\bar{z}_i)$ corresponds to a boundary point of the geodesic $(\chi,\psi)$, which implies $x_i=\bar{x_i}$, while the other is the intersection point of $(\chi_\Sigma,\psi_\Sigma)$ and $(\chi,\psi)$. 
Note that $\eta$ is divergent for all contributing geodesics and so we must regularize \eqref{eq: geodesic distance poincare} before using it in \eqref{eq: C i.t.o eta}.

Rather than working in terms of the cross-ratio, we may express \eqref{eq: C i.t.o eta} in terms of another function of $\eta$.
One interesting approach is to write $\eta$ in terms of well-known quantities from information theory. 
In particular, we may express $\eta$ in terms of entanglement entropies,
\begin{equation}
\label{eq: length of geodesic segment}
	\eta
	= 
	e^{\frac{3}{c}\kappa}+1\ ,
\end{equation}
where
\begin{equation}
	\kappa
	=
	S([x_1,x_2])+S([u,v])
	-S([x_2,v])-S([u,x_1])\ .
\end{equation}
The mutual information $I$ of two intervals $A$ and $B$ is defined as $I(A,B)=S(A)+S(B)-S(AB)$.
When $(x_1-u)(v-x_2)$ is sufficiently small, $\kappa$ coincides with the mutual information
\begin{equation}
\label{eq: kappa is a mutual information}
	\kappa
	=
	I([x_1,x_2],[u,v]^c)\,.
\end{equation}
For type (b) and (c) geodesics, as the cutoff is brought to zero the product $(x_1-u)(v-x_2)\to 0$ as well, guaranteeing that the interpretation of $\kappa$ as mutual information \eqref{eq: kappa is a mutual information} is valid.
Writing \eqref{eq: C i.t.o eta} in terms of $S$ and $\kappa$ gives
\begin{equation}
\label{eq: C i.t.o mi}
	\mathcal{C}(\chi_\Sigma,\psi_\Sigma)
	=
	- \frac{9}{\pi c^2}
	\int_{\text{type}\,b}d\chi d\psi S\partial_\psi^2S
	- \frac{3}{2\pi c}
	\int_{\text{type}\,c}d\chi d\psi \log\Big(e^{\frac{3}{c}\kappa}+1\Big)\partial_\psi^2 S\ .
\end{equation}
We note that \eqref{eq: C i.t.o mi} involves a single integral over kinematic space, as opposed to the double integral of \eqref{eq: complexity i.t.o. CFT quantities}. 
We stress, however, that the derivation of \eqref{eq: C i.t.o mi} was based on the identification of bulk chords with pairs of boundary intervals, which not only requires explicit information from the bulk, but is special to the vacuum.
Nevertheless, \eqref{eq: C i.t.o mi} may offer clues to the interpretation bulk volumes within field theory. 
Finally, we comment that, although we expect a similar relation to hold in global $\ads_3$, we have not worked it out explicitly.

%%%%%%%%%%%%%%%%%%%%%%%%%%%%%%%%%%%%%%%%%%%%%%%%%%%%%%%%%%%%%%%%%%%%%%%%%%%%%%%%
\section{Excited States}
\label{sec: excited states}
%%%%%%%%%%%%%%%%%%%%%%%%%%%%%%%%%%%%%%%%%%%%%%%%%%%%%%%%%%%%%%%%%%%%%%%%%%%%%%%%
So far, we have considered the volume formula only for vacuum $\ads_3$. 
However, the same tools can be applied to any geometry that is a quotient of $\ads_3$ by a discrete group of isometries. 
This is possible because the kinematic spaces for these geometries are themselves a quotient of the $\ads_3$ kinematic space.

We focus on the \emph{conical defect} and (static) \emph{BTZ black hole} geometries. 
In the CFT, these correspond to light primary excitations and finite temperature states, respectively.
Because the kinematic spaces of these geometries are quotients of the vacuum kinematic space, it follows that the volume formula derived above for vacuum AdS still applies, with the measure $\omega$ inherited from the quotienting procedure.

Before we examine the volume formula in detail, let us point out several important differences with the vacuum case. 
These differences stem from the fact that a given boundary interval may now have multiple geodesics terminating on its endpoints. 
The RT formula implies that only for one of these -- the shortest, or \emph{minimal}, geodesic -- does its length correspond to the entanglement entropy. 
Non-minimal geodesics come in two classes.
The first are those anchored at the endpoints of a boundary interval, but are not minimal; these we call \emph{winding geodesics}.
The second are those with only one endpoint lying on the boundary. 

In general, the bulk contains regions that intersect \emph{no} minimal geodesic. 
Such regions untouched by entanglement entropy go by the name \textit{entanglement shadow}. 
Hearteningly, the entanglement shadow is probed by non-minimal geodesics, which are naturally included as members of the quotient kinematic spaces.
In the literature, non-minimal geodesics constitute the building blocks of an observable called \textit{entwinement} \cite{Balasub14}, and were conjectured to measure correlations between internal degrees of freedom. 
For symmetric orbifold theories, an expression for entwinement with the correct properties was proposed in \cite{Balasub16}.

The non-uniqueness of geodesics implies that $\omega$ is no longer given simply in terms of the entanglement entropy, a consequence of the fact that at large $c$ the entropy is sensitive only to the shortest geodesic. 
In order to express the subregion complexity in terms of CFT quantities, we would therefore need to compute the lengths of non-minimal geodesics by alternate means, something that remains impossible with the present tools. 
Lastly, in contrast to the conical defect, thermal states also possess geodesics that pass through the black hole horizon. 
We expect the contributions from such geodesics to be associated to the thermal part of the reduced density matrix, as we will discuss below.

We begin this section by studying volumes first in conical defect geometries, followed by the BTZ black hole. 
We end by examining the decomposition of subregion complexity into contributions from entanglement entropy and from non-minimal geodesics, and a discussion of its physical significance.

%%%%%%%%%%%%%%%%%%%%%%%%%%%%%%%%%%%%%%%%%%%%%%%%%%%%%%%%%%%%%%%%%%%%%%%%%%%%%%%%
\subsection{Primary States: The Conical Defect $\mathcal{CD}_N$}
\label{sec: conical defect}
\begin{figure}[t]
\begin{center}
\includegraphics[scale=0.22]{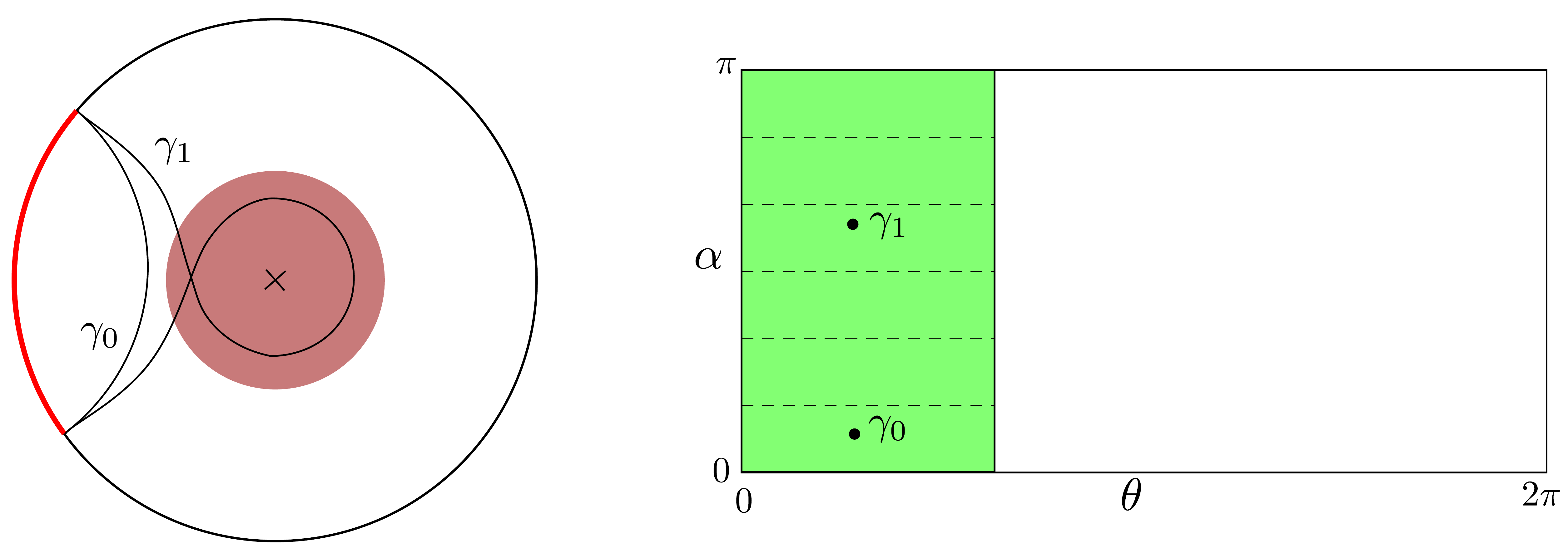} 
\end{center}
\caption{Left: the conical defect for the case $N=3$. A minimal ($\gamma_0$) and a non-minimal ($\gamma_1$) geodesic -- reaching into the entanglement shadow (red) -- are depicted. 
	Right: kinematic space for $N=3$. 
	Minimal geodesics correspond to the lowest and upmost $(n=0)$ sectors. The rest correspond to entwinements such as $\gamma_1$.}
\label{fig: windings}
\end{figure}

%%%%%%%%%%%%%%%%%%%%%%%%%%%%%%%%%%%%%%%%%%%%%%%%%%%%%%%%%%%%%%%%%%%%%%%%%%%%%%%%
The metric of the conical defect geometry $\mathcal{CD}_N$ takes the same form as the AdS$_3$ geometry \eqref{eq: metric global AdS3}, except that the periodicity of $\phi$ is modified to $\phi\sim\phi+2\pi/N$ $(N\in\mathds{N})$. 
More concretely, it can be thought of as a quotient of pure AdS$_3$,
\be
\mathcal{CD}_N=\frac{\textrm{AdS}_3}{\mathds{Z}_N} \,.
\ee
The kinematic space metric of the conical defect, as worked out in \cite{Cresswell:2017mbk}, takes the same form  \eqref{eq: ds and omega for symmetric spaces} as in the vacuum,
\begin{equation}
	ds^2_\cK
	=
	-\frac{1}{2}\partial^2_\alpha S(-d\alpha^2+d\theta^2)\,,
	\quad
	\omega
	=
	-\frac{1}{2}\partial^2_\alpha Sd\theta\wedge d\alpha \,,
\end{equation}
the difference being that now $\theta\sim\theta+2\pi/N$.
As a result, some geodesics have lengths computed by entanglement entropy, while others are non-minimal geodesics, possibly winding multiple times around the singularity. 
The fundamental region is divided into sectors $\alpha\in \cW_n^\pm$, where
\be
\cW^+_n = \biggl(\frac{n\pi}{2N},\,\frac{(n+1)\pi}{2N}\biggr],
\qquad
\cW^-_n = \biggl[\frac{(2N-n-1)\pi}{2N},\,\frac{(2N-n)\pi}{2N}\biggr),
\ee
with $n\in\{0,...,N-1\}$. 
$\cW_n^\pm$ describes the geodesics with winding number $n$ and orientation $\pm$.
In particular, $n=0$ corresponds to minimal geodesics, while geodesics with $n\ne 0$ are non-minimal. 
An illustration of these sectors is given in \figref{fig: windings} for the case $N=3$.

%--------------------------------------------
%--------------------------------------------

Including non-minimal geodesics is not only necessary, but also suffices to compute the volume of the constant time slice of the conical defect, 
\begin{align}
\label{eq: complexity CD}
 \cC(\mathcal{CD}_N)&=\frac{9}{8\pi c^2}\underbrace{\int_0^{2\pi/N}d\theta}_{2\pi/N}
		   \underbrace{\int_\xi^{\pi-\xi} d\alpha\int_{\Delta^\xi_{pp'}}d\theta'd\alpha'\partial_\alpha^2S(\alpha)
		   \partial_{\alpha'}^2S(\alpha')}_{\textrm{cf. \eqref{eq: complexity circle}}}
		 =\frac{1}{N}\cC(\text{circle})\, .
\end{align}
Dropping contributions from non-minimal geodesics, on the other hand, leads to expressions with the wrong divergence structure. 
For example, if we evaluate the outer integral in \eqref{eq: complexity CD} over minimal geodesics alone, we obtain
\begin{align}
 -\frac{1}{2}&\int_0^{\frac{2\pi}{N}}d\theta\biggl[\int_\xi^{\frac{\pi}{2N}}d\alpha+\int^{\pi-\xi}_{\frac{(2N-1)\pi}{2N}}d\alpha\biggr]
		\underbrace{\int_{\Delta^\xi_{pp'}}d\theta'd\alpha'\partial_{\alpha'}^2S(\alpha')}_{\textrm{cf. \eqref{eq: ent entropy from kin sp}}}\,\partial_\alpha^2S(\alpha)\notag\\
		=&\frac{\vol(\textrm{AdS}_3)}{N}-\frac{\pi}{N}\biggl(2\cot\biggl(\frac{\pi}{2N}\biggr)\log\biggl(\frac{\sin(\pi/2N)}{\xi}\biggr)-\frac{\pi}{N}(N-1)\biggr)+\cO(\xi^2)\, .
\label{eq:CD minimal geodesics only}
\end{align}
Here, $\vol(\textrm{AdS}_3)$ is the volume \eqref{eq: complexity circle result} of a constant time slice of AdS$_3$. 
Only for the vacuum ($N=1$) does this coincide with \eqref{eq: complexity CD}. 
In fact, away from $N=1$ the logarithmic dependence on the cutoff is not even consistent with a volume in an asymptotically $\ads_3$ spacetime, which should exhibit as its sole singularity a term scaling as $\xi^{-1}$ \cite{Abt:2017pmf}.
Of course, the problematic logarithm of \eqref{eq:CD minimal geodesics only} drops out when we include non-minimal geodesics.

Finally, we emphasize that non-minimal geodesics are required not only to compute volumes in the entanglement shadow, but also for regions outside of it, as is evident from \figref{fig: windings}.

%%%%%%%%%%%%%%%%%%%%%%%%%%%%%%%%%%%%%%%%%%%%%%%%%%%%%%%%%%%%%%%%%%%%%%%%%%%%%%%%
\subsection{Subregion Complexity at Finite Temperatures}
\label{sec: subregion complexity at finite temperatures}
%%%%%%%%%%%%%%%%%%%%%%%%%%%%%%%%%%%%%%%%%%%%%%%%%%%%%%%%%%%%%%%%%%%%%%%%%%%%%%%%
We now turn to the volume formula in BTZ black hole geometries \cite{Banados:1992wn}. 
We restrict ourselves for simplicity to the spinless solution ($J=0$), whose metric is
\be
	ds^2 = -\frac{r^2-r_0^2}{L^2}dt^2 + \frac{L^2}{r^2-r_0^2}dr^2 + r^2d\phi^2 \,,
	\qquad
	\phi\sim\phi+2\pi\,.
\label{eq:BTZ}
\ee
Our discussion begins with a brief description of BTZ kinematic space%
\footnote{Two versions of BTZ kinematic space have appeared in the literature: quotient kinematic spaces of the type used here also appeared in \cite{Zhang:2016evx, Czech:2015kbp}, whereas the kinematic space of \cite{Asplund:2016koz} contained only minimal geodesics.}
and the generalization of the volume formula \req{eq:volume formula} to it.
We then compute the BTZ subregion complexity using this formula, written in terms of the Poincar\'e patch measure of \secref{sec:pp volume}.

%%%%%%%%%%%%%%%%%%%%%%%%%%%%%%%%%%%%%%%%%%%%%%%%%%%%%%%%%%%%%%%%%%%%%%%%%%%%%%%%
\subsubsection*{Kinematic Space of the BTZ Black Hole}
%%%%%%%%%%%%%%%%%%%%%%%%%%%%%%%%%%%%%%%%%%%%%%%%%%%%%%%%%%%%%%%%%%%%%%%%%%%%%%%%
The BTZ black hole geometry \req{eq:BTZ} is obtained from $\ads_3$ by quotienting by a discrete group of isometries with a particularly simple form in Poincar\'e patch coordinates.
Writing the Poincar\'e patch metric in the form
\be
ds^2 = L^2\frac{-(dx^0)^2+(dx^1)^2 + dz^2}{z^2} = L^2\frac{dx^+dx^- + dz^2}{z^2} \,,
\ee
with $x^\pm = x^1 \pm x^0$, the map
\be
x^\pm = \Bigl(1-\frac{r_0^2}{r^2}\Bigr)^{1/2} e^{\frac{r_0}{L}(\phi\pm t/L)} \,,
\qquad
z = \frac{r_0}{r} e^{\frac{r_0}{L}\phi} \,,
\ee
is a local isometry to \req{eq:BTZ}.
The periodicity $\phi\sim\phi+2\pi$ of the BTZ coordinates requires us to identify the points
\be
(x^0,x^1,z) \sim e^{2\pi r_0/L}(x^0,x^1,z) \,.
\label{eq:pp identification}
\ee
This identification generates a group of infinite order, and the quotient of the Poincar\'e patch by it is isometric to a region in the maximally extended BTZ geometry of mass $M=r_0^2/L^2$. 

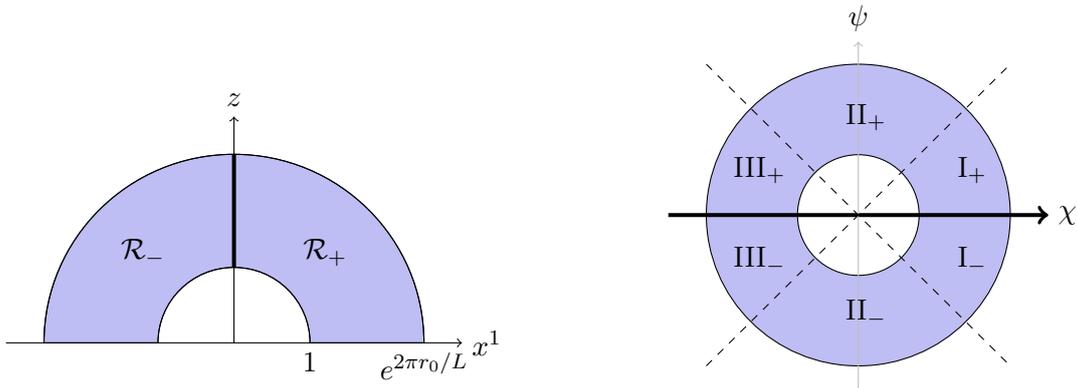
\begin{figure}
\centering
 \newcommand{\inr}{1}
 \newcommand{\outr}{2.5}
  \begin{tikzpicture}
      \filldraw [fill=\figblue] (\outr,0) node [below] {$e^{2\pi r_0/L}$} arc(0:180:\outr) -- (-\inr,0) arc(180:0:\inr) node [below] {1}  -- (\outr,0) ;
      \draw (\inr,0) arc (0:180:\inr) ;
      \draw (\outr,0) arc (0:180:\outr) ;
      \draw [->] (-3,0) -- (3,0) node [right] {$x^1$} ;
      \draw [->] (0,0) -- (0,3) node [above] {$z$} ;
      \draw [line width=1.5pt] (0,\inr) -- (0,\outr) ;
      \node at (1.2,1.2) {$\cR_+$} ;
      \node at (-1.2,1.2) {$\cR_-$} ;     
  \end{tikzpicture}
 \hspace{10ex}
  \newcommand{\sz}{2}
  \begin{tikzpicture}
   \path [draw,fill=\figblue,even odd rule] (0,0) circle(.8) (0,0) circle(2) ;
   \draw [->,line width=1.5pt] (-2.5,0) -- (2.5,0) ;
   \node at (2.5,0) [right] {$\chi$} ;
   \draw [->,gray!50] (0,-2.3) -- (0,2.3) ;
   \node at (0,2.3) [above] {$\psi$} ;
   \draw [dashed] (-\sz,-\sz) -- (\sz,\sz) ;
   \draw [dashed] (-\sz,\sz) -- (\sz,-\sz) ;
   \node at (1.5,0.6) {I$_+$} ;
   \node at (1.5,-0.6) {I$_-$} ;
   \node at (0.1,1.3) {II$_+$} ;
   \node at (0.1,-1.3) {II$_-$} ;
   \node at (-1.3,0.6) {III$_+$} ;
   \node at (-1.3,-0.6) {III$_-$} ;
  \end{tikzpicture}
\caption{%
Figure on left: Fundamental region for the spatial slice of the 2-sided black hole in Poincar\'e patch coordinates \eqref{eq:BTZ K coord I}. The horizon (dark line) separates the two asymptotic regions $\cR_+$ and $\cR_-$. 
Figure on right: The six fundamental domains for kinematic space in $(\chi,\psi)$ coordinates.
The ratio of the outer and inner radii is $e^{2\pi r_0/L}$.
The metric diverges as one approaches the dark line $\psi=0$.}
\label{fig:pp quotient}
\end{figure}

\medskip
Because \req{eq:pp identification} preserves the locus $x^0=0$, the spatial slice $t=0$ of the black hole solution is the image of the spatial slice $x^0=0$ of the Poincar\'e patch,
\be
ds^2 = L^2\frac{(dx^1)^2 + dz^2}{z^2} \,.
\ee
The quotient space of this slice is, in fact, globally equivalent to the spatial slice of the two-sided BTZ black hole.
The quotient has the convenient fundamental domain
\be
1 \le (x^1)^2 + z^2 < e^{4\pi r_0/L}
\label{eq:fundamental domain}
\ee
(\figref{fig:pp quotient}).
Geodesics in the slice are mapped to geodesics, modulo the identification \req{eq:pp identification} acting simultaneously on both endpoints.
In other words, the kinematic space of BTZ is a quotient of the kinematic space of $\ads_3$.
As in section~\ref{sec:pp volume}, spatial geodesics in the Poincar\'e patch ending at $x^1=u,v$ can be written as $u=\chi-\psi$, $v=\chi+\psi$, giving kinematic space as the quotient manifold
\be
ds^2_{\cK_\text{BTZ}} = \frac{c}{6} \frac{d\chi^2-d\psi^2}{\psi^2} \,,
\qquad
(\chi,\psi) \sim e^{2\pi r_0/L} (\chi,\psi) \,.
\ee
Note that the horizon corresponds to the line $x^0=x^1=0$ in the Poincar\'e patch geometry.
The two sides of the black hole are separated by the horizon (\figref{fig:pp quotient}).

The points of BTZ kinematic space naturally break into six families.
In terms of the covering space coordinates $(u,v)$, they correspond to those with $0<u<v$ (region I$_+$); those with $u<0<v$ (region II$_+$); those with $u<v<0$ (region III$_+$); and the orientation reversal ($u\leftrightarrow v$) of these three sets.
Each of these regions has convenient coordinate systems.
For example, consider region I$_+$ the geodesics contained entirely in the positive asymptotic region with $0<u<v$ (\figref{fig:pp quotient}).
Setting
\be
v = e^{\frac{r_0}{L}(\theta+\alpha)} \,,
\qquad
u = e^{\frac{r_0}{L}(\theta-\alpha)} \,,
\label{eq:BTZ K coord I}
\ee
we obtain 
\be
ds^2_{\text{I}_+} = \frac{r_0^2}{L} \frac{d\theta^2-d\alpha^2}{\sinh^2(\frac{r_0\alpha}{L})} \,,
\qquad
\theta \sim \theta+2\pi\,,\;
\alpha\in\bR \,.
\ee
Both $ds^2_{\text{I}_+}$ and the vacuum kinematic space metric \eqref{eq:vacuum kinematic space} take the same form in the limit $\alpha\to 0$, as they should.
Geodesics are split into sectors given by $\alpha\in \cV_n$, 
\be
\cV_n = [2\pi n,2\pi(n+1)) \,.
\ee
Sector $\cV_n$ is said to have winding number $n$.
Similarly, for geodesics passing through the horizon ($u<0<v$) we may set 
\be
v = e^{\frac{r_0}{L}(\tilde\theta+\tilde\alpha)}
\qquad
u = -e^{\frac{r_0}{L}(\tilde\theta-\tilde\alpha)} \,,
\label{eq:BTZ K coord II}
\ee
leading to the geometry
\be
ds^2_{\text{II}_+} = \frac{r_0^2}{L}\frac{d\tilde\alpha^2-d\tilde\theta^2}{\sinh^2(\frac{r_0\tilde\alpha}{L})} \,,
\qquad
\tilde\theta \sim \tilde\theta+2\pi\,,\;
\tilde\alpha\in\bR \,.
\ee
The other four patches are related to these two by sign changes.
I$_+$, II$_+$, and III$_+$ all meet at a cuspoidal point, the (positively oriented) horizon geodesic, which corresponds in the two coordinate systems above to $\alpha\to\infty$ and $\tilde\alpha\to\infty$, respectively.

%%%%%%%%%%%%%%%%%%%%%%%%%%%%%%%%%%%%%%%%%%%%%%%%%%%%%%%%%%%%%%%%%%%%%%%%%%%%%%%%
\subsubsection*{Volume Formula at Finite Temperature}
%%%%%%%%%%%%%%%%%%%%%%%%%%%%%%%%%%%%%%%%%%%%%%%%%%%%%%%%%%%%%%%%%%%%%%%%%%%%%%%%
Our goal is to evaluate volumes in the BTZ black hole using the volume formula \req{eq:volume formula}.
Using the quotient construction, it is straightforward to apply the volume formula: given a volume in BTZ, we lift it to the fundamental domain \req{eq:fundamental domain} and apply \req{eq:volume formula}. 
Note that, for a region $Q$ lying entirely outside the horizon ($u,v>0$), \req{eq:volume formula} necessarily includes contributions not only from geodesics lying outside the black hole, but also from those passing through the horizon.
Pulling the resulting quantities back to BTZ kinematic space, the volume becomes
\be
\label{eq: volBTZ}
\frac{\vol(Q)}{4G_N} = \frac{1}{2\pi}\sum_{\cD}\int \lambda_Q \omega_\cD
\,,
\ee
where $\cD$ runs over the domains I$_\pm$, II$_\pm$, and III$_\pm$ of \figref{fig:pp quotient}. 
Obviously, the III$_\pm$ contribution vanishes when $Q$ is outside the horizon.
In the coordinates \eqref{eq:BTZ K coord I} and \eqref{eq:BTZ K coord II}, the Crofton form is
\be
\omega_\text{I,III} = \frac{c}{6}\frac{d\theta\wedge d\alpha}{\sinh^2(\frac{r_0}{L}\alpha)}
\qquad
\omega_\text{II} = -\frac{c}{6}\frac{d\tilde\theta\wedge d\tilde\alpha}{\sinh^2(\frac{r_0}{L}\alpha)} \,.
\ee
We are free to omit the ``$-$'' regions provided we multiply by an overall factor of 2.

In practice, the simplest way to perform computations is to work directly with a fundamental region in Poincar\'e patch. 
We now turn to the application of this method to evaluating the holographic subregion complexity in BTZ.

%%%%%%%%%%%%%%%%%%%%%%%%%%%%%%%%%%%%%%%%%%%%%%%%%%%%%%%%%%%%%%%%%%%%%%%%%%%%%%%%
\subsubsection*{Subregion Complexity and the Phase Transition}
%%%%%%%%%%%%%%%%%%%%%%%%%%%%%%%%%%%%%%%%%%%%%%%%%%%%%%%%%%%%%%%%%%%%%%%%%%%%%%%%
%-------------------------------------------
\begin{figure}[t]
\begin{center}
\includegraphics[scale=0.13]{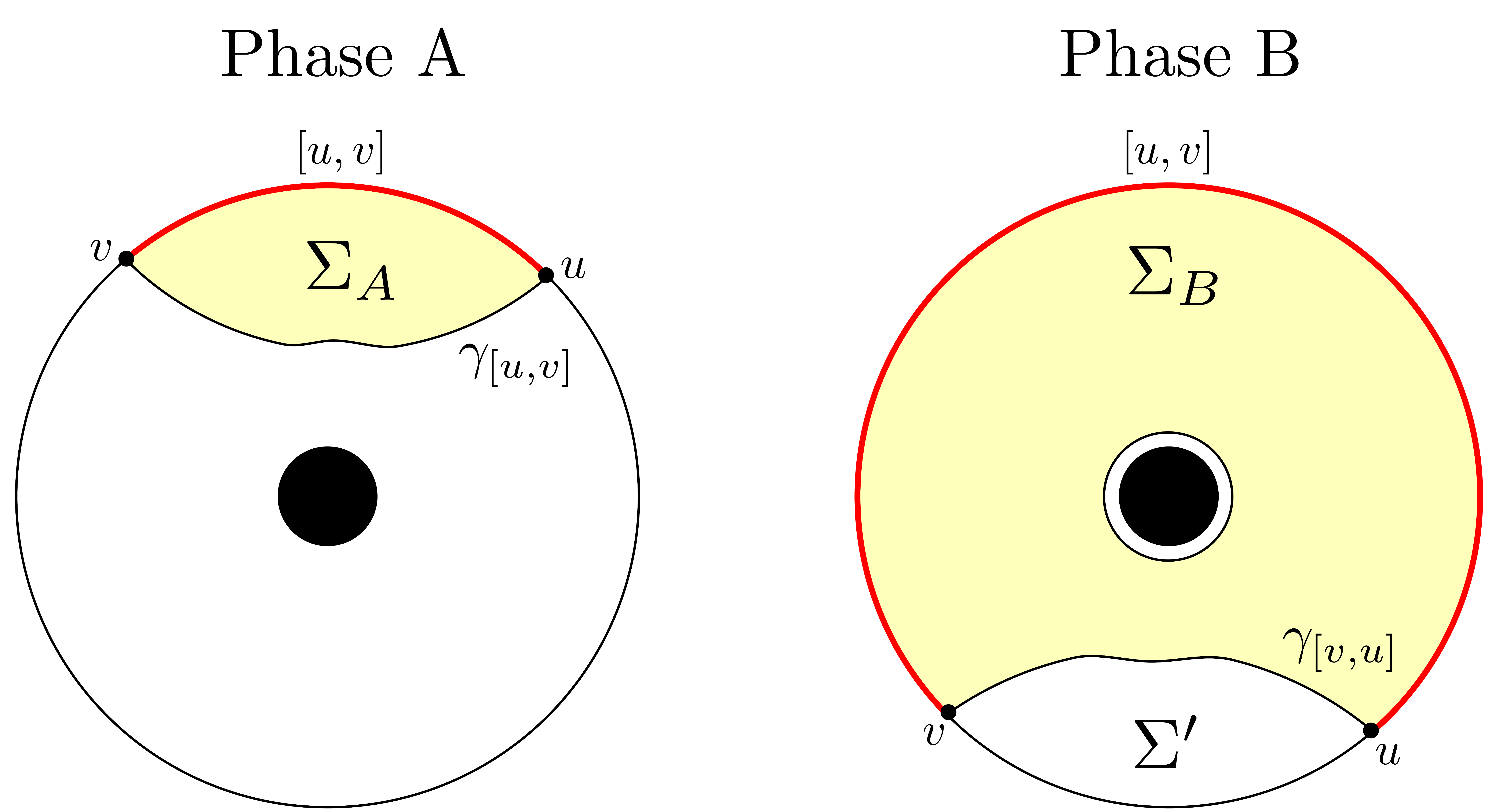} 
\end{center}
\caption{For the BTZ black hole, the RT surface undergoes a phase transition. If the entangling interval $[u,v]$ is too large, the RT surface
            is no longer the geodesic $\gamma_{[u,v]}$ lying on the same side of the black hole as $[u,v]$ (Phase A) but consists of the
            black hole horizon and the geodesic $\gamma_{[v,u]}$ lying on the opposite side of the black hole (Phase B). At the point of the
            phase transition the volume below the RT surface jumps from $\vol(\Sigma_A)$ to $\vol(\Sigma_B)$.}
\label{fig:phases AB}
\end{figure}
%-------------------------------------------
The holographic subregion complexity for an interval $(u,v)$ is the volume lying under the minimal curve homologous to that interval \cite{Hubeny13}. 
Depending on the size of the interval, there are two such curves, corresponding to distinct phases A and B (see \figref{fig:phases AB}). 
In phase A, the minimal curve is the curve in $\cV_0$ representing $(u,v)$.
In phase B, it is the union of the curve in $\cV_{-1}$ representing $(v,u)$ and the horizon geodesic. 
The dominant phase is the one whose curve has the shortest length.
We saw in \cite{Abt:2017pmf} that under the transition from phase A to phase B, the topological complexity \eqref{eq: topological complexity} increases by $2\pi$.

We compute the subregion complexity by applying the Poincar\'e patch volume formula to a fundamental region.
For simplicity, we utilize the bulk cutoff regularization. 
The correct domain of integration depends on the cutoff surface, which differs from that in the Poincar\'e patch.
The cutoff in the BTZ geometry lies at $r_\epsilon=Ll_\cft/\epsilon$, corresponding in the Poincar\'e patch to the $x^1$-dependent cutoff
\be
\epsilon(x) = \Bigl(\frac{r_\epsilon}{r_0}-1\Bigr)^{-1/2}|x^1| \,.
\ee

In phase A, except for the modified cutoff the computation now takes the same form as that of section~\ref{sec:pp volume}.
Integrating explicitly gives the final result
\begin{equation}
  \vol(\Sigma_A)=L^2\biggl(\Bigl(\frac{r_\epsilon}{r_0}-1\Bigr)^{1/2}\log\left(\frac{v}{u}\right)-\pi\biggr)+\mathcal{O}(\epsilon)=L^2\left(\frac{x}{\epsilon}-\pi\right)+\mathcal{O}(\epsilon) \,,
\label{eq:phase A volume}
\end{equation}
where $x=2\lcft\alpha$ is the length of the entangling interval in BTZ coordinates. 

In phase B, the integration region of the volume formula 
stretches from the outside of the
complementary geodesic up to the black hole horizon. 
Hence, the volume
can be computed by calculating the volume between boundary and horizon, and
then subtracting the volume that lies in the geodesic corresponding to the complementary interval $[v,u]$,
\begin{equation}
  \vol(\Sigma_B)=\vol(\textrm{outside horizon})-\vol(\Sigma')=L^2\left(\frac{x}{\epsilon}+\pi\right) \,,
\label{eq:phase B volume}
\end{equation}
where $\Sigma'$ is region under $\gamma_{[v,u]}$ (see \figref{fig:phases AB}) and $x$ is the length of the boundary interval. 
The volume of the outside horizon region is computed in kinematic space by taking the integral over all geodesics, cut off at the horizon for those that fall into the black hole, weighted with the Crofton form.
Comparing with \eqref{eq:phase A volume} gives the expected jump in complexity of $2\pi$. 

One can also do the calculation using a kinematic space cutoff similar to section~\ref{sec:pp volume}, although in that case the regularization scheme no longer matches that used in the gravitational calculation.

%%%%%%%%%%%%%%%%%%%%%%%%%%%%%%%%%%%%%%%%%%%%%%%%%%%%%%%%%%%%%%%%%%%%%%%%%%%%%%%%
\subsection{A Bound on Subregion Complexity from Entanglement Entropy}
\label{sec: fine structure}
%%%%%%%%%%%%%%%%%%%%%%%%%%%%%%%%%%%%%%%%%%%%%%%%%%%%%%%%%%%%%%%%%%%%%%%%%%%%%%%%
The discussion above makes it clear that, unlike the vacuum case (\secref{sec: vacuum subregion complexity}), the subregion complexity of quotient geometries depends not only on entanglement entropies, but also receives contributions from non-minimal geodesics. 
It is still possible, however, to isolate contributions to subregion complexity depending only on entanglement entropies. 

We first consider the conical defect geometry.
Here, subregion complexity gets contributions from winding geodesics, which spoils the one-to-one correspondence between geodesics and entangling intervals. 
Those geodesics are nonetheless still attached to pairs of boundary points, and are thus associated to entangling intervals in a natural way. 
We can therefore organize the expression \eqref{eq: complexity i.t.o. CFT quantities} for the subregion complexity of a boundary interval $A$ in the form
\begin{equation}
\label{eq: C interpretation CD}
	\cC(A)
	=
	\int d\hat\theta\, d\hat\alpha\Big(
		F^{CD}_A
		+
		G^{CD}_A\Big)\,,
\end{equation}
where $\hat{\theta}$ and $\hat{\alpha}$ parametrize the set of boundary intervals as in the vacuum kinematic space \eqref{eq:vacuum kinematic space}.
$F^{CD}_A$ denotes the part of subregion complexity containing only entanglement entropies (that is, the subregions of $\cK$ and $\Delta_{pp'}$ in the integral expression \eqref{eq: complexity i.t.o. CFT quantities} due to minimal geodesics).
The term $G^{CD}_A$ contains the remaining contributions from non-minimal geodesics winding around the singularity.

As with the conical defect, the subregion complexity of the BTZ black hole receives contributions from winding geodesics.
However, black holes have a new class of geodesics contributing to subregion complexity: those that pass through the black hole horizon $r=r_0$. 
Because they are associated to finite temperatures, we refer to them as `thermal contributions'. 
The subregion complexity now takes the form 
\begin{equation}
\label{eq: C interpretation BTZ}
	\cC(A)
	=
	\int d\hat\theta d\hat\alpha\Big(F_A^{BTZ}+G_{A}^{BTZ}\Big)
	\,+\,
	\text{thermal contributions}\,,
\end{equation}
where $F^{BTZ}_A$ denotes those contributions from entanglement entropies alone, and $G_{A}^{BTZ}$ represents contributions of winding geodesics. 
Because the thermal contributions have only one endpoint on the boundary, they cannot be associated to entangling intervals, and so cannot be included in the integral over entangling intervals of \eqref{eq: C interpretation BTZ}. 

Although the pure entanglement contributions in these geometries do not suffice to compute the subregion complexity, we note that because the other contributions are all positive, the pure entanglement contribution $\int F$ does place a lower bound on holographic subregion complexity.

This situation bears a certain resemblance to the distinction between spectrum complexity and basis complexity of \cite{Agon:2018zso}. 
We can think of the part of holographic subregion complexity containing only entanglement entropies as analogous to spectrum complexity (since entanglement entropy only depends on the spectrum of a state), while the remaining contributions build up the more detailed basis complexity. 
However, we should stress that the entanglement contributions contain information not only about the spectrum of the reduced density matrix of $A$ (i.e., entanglement entropy), but also about the spectra of all intervals overlapping $A$.

%%%%%%%%%%%%%%%%%%%%%%%%%%%%%%%%%%%%%%%%%%%%%%%%%%%%%%%%%%%%%%%%%%%%%%%%%%%%%%%%
\section{Discussion}
\label{sec: discussion}
%%%%%%%%%%%%%%%%%%%%%%%%%%%%%%%%%%%%%%%%%%%%%%%%%%%%%%%%%%%%%%%%%%%%%%%%%%%%%%%%
In this paper we studied volumes in a fixed spatial slice of the $\ads_3$ vacuum of a gravitational theory with a holographic dual CFT. 
Our primary technical result is an expression for the volume of any region in that slice as an integral over the kinematic space of the dual CFT (\secref{sec: volume formula proof}), a formula we applied to express the volume of the region contained under a geodesic in terms of entanglement entropies alone (\secref{sec: vacuum subregion complexity}). 
Following the proposal of \citep{Alishahiha:2015rta}, we refer to this quantity as `subregion complexity'.
The volume formula is a manifestation of the relation between entanglement in a QFT and the geometry of its bulk dual, as captured by the motto `entanglement builds geometry' \cite{VanRaamsdonk:2010pw}. 
Because our result represents this volume purely in terms of CFT quantities, it may help provide insight into the significance of this quantity for CFT.

The description of locally $\ads_3$ geometries as quotient spaces of the vacuum by a discrete group of isometries allowed us to extend our primary result \eqref{eq: complexity i.t.o. CFT quantities} to express subregion complexities for primary excitations (conical defect of \secref{sec: conical defect}) and thermal states (BTZ black hole of \secref{sec: subregion complexity at finite temperatures}) as integrals over appropriate kinematic spaces. 
In these cases, however, it was necessary to include not only the minimal geodesics corresponding to entanglement entropies, but also contributions from non-minimal geodesics related to entwinement and to the presence of a horizon.

While equation \eqref{eq: complexity i.t.o. CFT quantities}, together with the prescription of \secref{sec: regions of integration for complexity}, computes the vacuum subregion complexity in terms of CFT quantities, it is potentially useful to reformulate this integral in terms of the correspondence \cite{Wen:2018whg} between bulk chords and nested pairs of boundary intervals. 
This correspondence let us express the chord length $\lambda_\Sigma$ appearing in the volume formula \eqref{eq:vol formula for Sigma} for the Poincar\'e patch
using the conformal cross-ratio of these two intervals (\secref{sec: chord lenghts for Pp and mi}). 
In many cases this can further be written in terms of a mutual information \eqref{eq: C i.t.o mi}. 
Such a reformulation may yield further insights into the volume formula.

\medskip
A major motivation for deriving bulk volumes from within field theory is the complexity $\!=\!$ \mbox{}volume conjecture \cite{Susskind:2014moa}, on which Alishahiha's proposal -- that the volume below an RT surface is a measure of the complexity of the corresponding reduced density matrix \citep{Alishahiha:2015rta} -- was based. 
This proposal is difficult to test: in contrast to entanglement entropy, there is as yet no entirely satisfactory notion for complexity in QFT, although some progress has been made toward such a notion for free QFTs \cite{Chapman:2017rqy, Jefferson:2017sdb, Hackl:2018ptj, Yang:2017nfn, Kim:2017qrq, Reynolds:2017jfs, Khan:2018rzm}.
Instead of considering its interpretation as a complexity, we focused on a complementary question: how can Alishahiha's bulk geometric quantity, the holographic subregion complexity, be computed within CFT? 
At least in the vacuum of a large-$N$ CFT, the volume formula provides an answer to this question. 

Let us now consider the implications of our results, assuming Alishahiha's proposal is valid. 
The first is that in the vacuum state, subregion complexity can be computed purely in terms of entanglement entropy, suggesting that vacuum subregion complexity is encoded in the spectrum of single-interval entanglement, at least in the large-$N$ limit. 
On the other hand, in non-vacuum geometries we found that volumes received contributions other than single-interval entanglement entropies. 
Nevertheless, a part of the complexity in each geometry we considered is determined by entanglement entropies alone, as expressed in
\eqref{eq: C interpretation CD} and \eqref{eq: C interpretation BTZ}. 
Therefore, if we restrict our formula to include contributions only from entanglement entropies, the resulting information-theoretic quantity constitutes a \emph{lower bound} for Alishahiha's holographic subregion complexity that is built entirely from single-interval entanglement entropies. 
The relation between pure entanglement contributions and subregion complexity also shares features with that between the spectrum and basis complexities of \cite{Agon:2018zso}. 

Conical defect geometries require us to supplement entanglement entropies of single intervals by non-minimal geodesics associated to single intervals, presumably reflecting the more involved structure of entanglement in excited states. 
The volume formula for the BTZ black hole requires us to include further objects, geodesics that enter the black hole horizon. 
Because these do not bound any boundary interval in the BTZ geometry or the two-sided black hole, they have no clear interpretation in terms of entanglement.
Such geodesics are, however, crucial to computing e.g. the length of the horizon in kinematic space, suggesting that their lengths contain important information about the thermal density matrix that the black hole geometry represents.
It is important to understand the role of these geodesics in greater detail.

A consequence of our construction 
%in terms of entanglement entropies (and for excited states also entwinement) 
is that holographic subregion complexity (in the sense of Alishahiha) in vacuum and thermal states is universal, in that it depends only on the central charge of the field theory. 
%Currently known complexity constructions for free field theories 
Field theory proposals for the complexity of Gaussian states do not possess this property, exhibiting for example a different behavior between bosons and fermions.
%since the complexity behavior for bosons and fermions differs even if it is calculated using the same method for both cases 
\cite{Jefferson:2017sdb,Hackl:2018ptj}.%
\footnote{We would like to thank the referee for pointing this out.}
%However our results only apply for interacting field theories in the large $N$ limitIt is important to note
We note, however, that our results only compute complexity in strongly interacting theories in the large $N$ limit, and only if the complexity $\!=\!$ \mbox{}volume conjecture holds true.
Nevertheless, this universality constitutes a strong test of the conjecture once a  field theory computation of complexity for such theories is known.

\medskip
A number of important questions regarding the volume formula, and its interpretation in terms of complexity, remain unanswered. 
One notable task is to generalize the volume formula to geometries with small \emph{local} variations away from vacuum $\ads_3$. 
Note that such geometries represent small $(\Delta E\ll c/\lcft)$ deviations from the vacuum state, so it is reasonable to expect that entwinement will play no role.
If the complexity $\!=\!$ volume conjecture is correct, such a formula may give important insights into the structure of complexity, as well as its relationship to entwinement.

It is also desirable to understand the relationship between subregion complexity and state complexity more deeply. 
Susskind's original complexity $\!=\!$ volume proposal \cite{Susskind:2014moa} was based on the expected features of time evolution in a two-sided eternal black hole background.
It would be very useful to study the expected behavior of subregion complexity in time-dependent quantum systems, and then to compare their qualitative behavior with gravitational computations. 

Another intriguing direction would be to study the relationship of our approach to that of \cite{Caputa:2017urj,Bhattacharyya:2018wym}, where complexity is defined in terms of a path integral optimization procedure. 
Optimizations have been related to kinematic space \cite{Czech:2017ryf}, and thus it would be interesting to understand how such approaches might be related to the concept of complexity as volumes of AdS regions and their computation via the volume formula.

%%%%%%%%%%%%%%%%%%%%%%%%%%%%%%%%%%%%%%%%%%%%%%%%%%%%%%%%%%%%%%%%%%%%%%%%%%%%%%%%
\begin{acknowledgments}
We thank Michal Heller, Haye Hinrichsen, Ren\'e Meyer, Rob Myers, Andy O'Bannon, Fernando Pastawski and Ignacio A. Reyes for helpful discussions.
The work of CMT is supported by an Alexander von Humboldt Research Fellowship. 
\end{acknowledgments}
%%%%%%%%%%%%%%%%%%%%%%%%%%%%%%%%%%%%%%%%%%%%%%%%%%%%%%%%%%%%%%%%%%%%%%%%%%%%%%%%

%%%%%%%%%%%%%%%%%%%%%%%%%%%%%%%%%%%%%%%%%%%%%%%%%%%%%%%%%%%%%%%%%%%%%%%%%%%%%%%%
\appendix
%%%%%%%%%%%%%%%%%%%%%%%%%%%%%%%%%%%%%%%%%%%%%%%%%%%%%%%%%%%%%%%%%%%%%%%%%%%%%%%%
\section{Subregion Complexity for the Semicircle}
\label{ap: subregion complexity for the semicircle}
In this section we show how to obtain the subregion complexity of the
semicircle in the vacuum state given by \eqref{eq: complexity for semicircle}
from equation \eqref{eq: complexity i.t.o. CFT quantities}. Even
though this calculation does not require the bulk at any point, we
will refer to it during the computation in order to explain certain
steps in the most intuitive way.

\begin{figure}
\begin{center}
\includegraphics[scale=0.3]{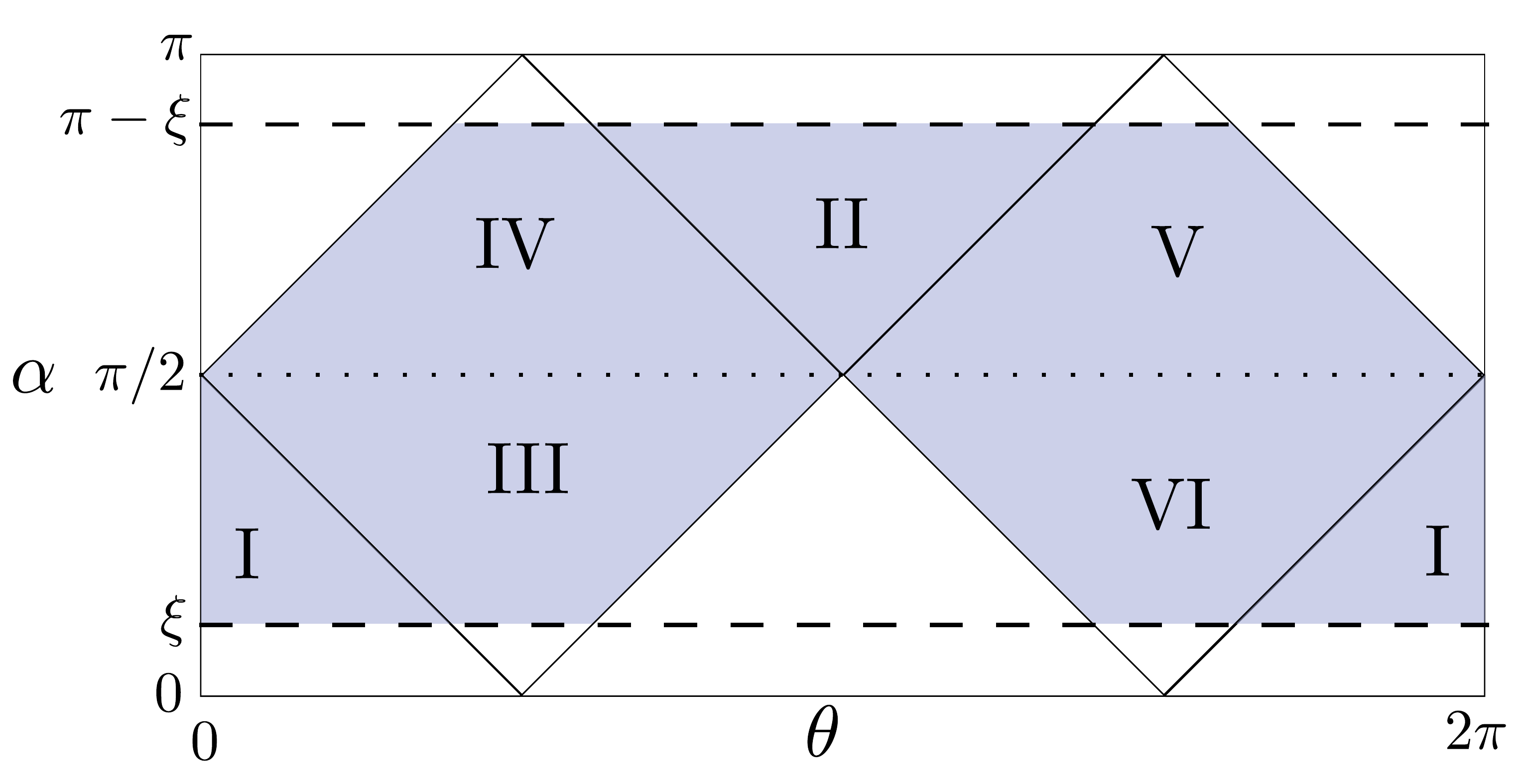} 
\end{center}
\caption{The set of type (b) and (c) intervals for the entangling interval
			$(0,\pi/2)$. I and II denote the intervals of type (b) while
			III-VI are the intervals of type (c). The cutoffs are set to
			$\alpha'=\xi$ and $\alpha'=\pi-\xi$. It is easy to verify that I
			and II contribute in the same way to the complexity. So do
			III-VI.}
\label{fig: region of int for semicircle}
\end{figure}

Without loss of generality, we assume that the semicircle is centered
around $0$, i.e. the boundary interval is
$(\theta_\Sigma=0,\alpha_\Sigma=\pi/2)$. To regularize the integrals, we introduce the usual cutoffs at $\alpha=\xi$ and $\alpha=\pi-\xi$.
Therefore we find from \eqref{eq: complexity i.t.o. CFT quantities}
\begin{equation}
	\cC(\text{semicircle})
	=
	\frac{9}{8\pi c^2}\int_0^{2\pi}d\theta\int_\xi^{\pi-\xi}d\alpha\,
		\Lambda_\xi(\theta,\alpha)\partial^2_\alpha S(\alpha)\,,
\end{equation}
where
\begin{equation}
\label{eq: cutoff chord length}
	\Lambda_\xi(\theta,\alpha)
	=
	\int_{\Delta^\xi_{pp'}}d\theta'd\alpha'
			 \partial_{\alpha'}^2S(\alpha')\,.
\end{equation}
Since the entangling intervals of type (a) do not contribute,
we only need to consider the ones of type (b) and (c). Those are depicted
in \figref{fig: region of int for semicircle}.
%--------------------------------------------
%--------------------------------------------
Due to the symmetry of the volume form, the regions I and II as well as III-VI give the same contribution to the complexity. Consequently $\cC$ is given by
\begin{equation}
  \begin{split}
	\cC(\text{semicircle})
	=
	\frac{9}{4\pi c^2}\Big(
	&
	\int_\xi^{\pi/2}d\alpha\int_{\alpha-\pi/2}^{\pi/2-\alpha}d\theta\,
		\Lambda_\xi(\theta,\alpha)\partial^2_\alpha S(\alpha)
	\\
	&
	+
	2\int_\xi^{\pi/2}d\alpha\int_{\pi/2-\alpha}^{\pi/2+\alpha}d\theta\,
		\Lambda_\xi(\theta,\alpha)\partial^2_\alpha S(\alpha)\Big)\,.
  \end{split}	
\end{equation}
The first integral computes the contribution from type (b) intervals (region I), while the second integral computes the contribution from type (c) intervals (region III). For the type (b) intervals $\Lambda_\xi$ is already known and given by \eqref{eq: ent entropy from kin sp}. So we obtain
\begin{equation}
\label{eq: calc complexity for semicircle alt}
\begin{split}
	\cC(\text{semicircle})
	=
	\frac{9}{4\pi c^2}\Big(
	&
	-\frac{8c}{3}
	\int_\xi^{\pi/2}d\alpha\int_{\alpha-\pi/2}^{\pi/2-\alpha}d\theta\left(
		\log\left(\frac{\sin(\alpha)}{\sin(\xi)}\right)
		+
		\xi\cot(\xi)\right)\partial^2_\alpha S(\alpha)
	\\
	&
	+
	2\int_\xi^{\pi/2}d\alpha\int_{\pi/2-\alpha}^{\pi/2+\alpha}d\theta\,
		\Lambda_\xi(\theta,\alpha)\partial^2_\alpha S(\alpha)\Big)\,.
\end{split}	
\end{equation}

\begin{figure}
\begin{center}
\includegraphics[scale=0.22]{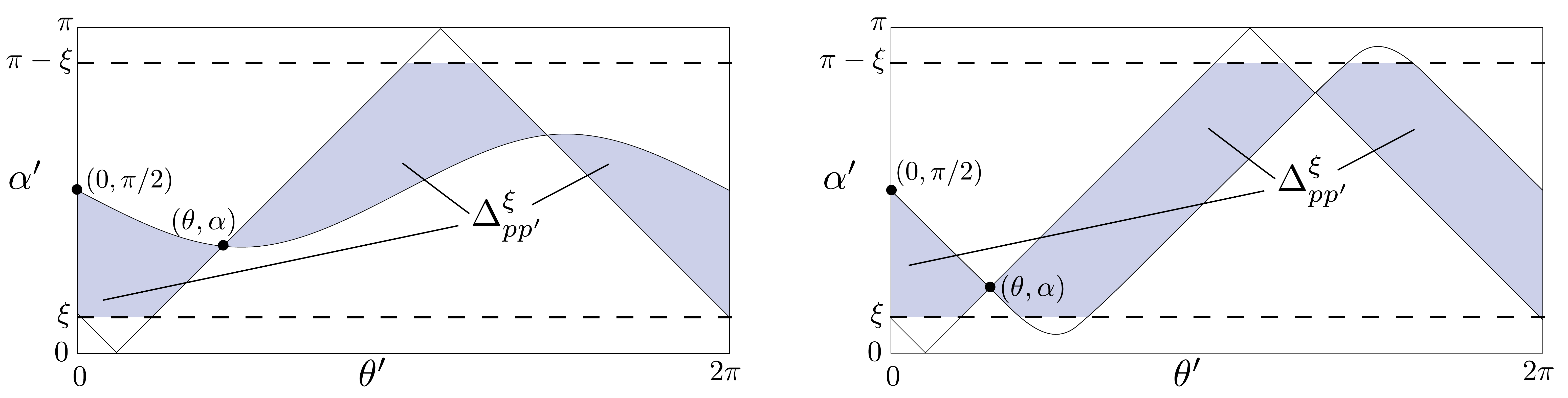} 
\end{center}
\caption{In the calculation of the complexity of the semicircle $(0,\pi/2)$,
			entangling intervals of type (c) show up. By construction the
			corresponding regions of integration are inter alia bounded by
			point curves. When we introduce cutoffs at $\alpha'=\xi$ and
			$\alpha'=\pi-\xi$ we obtain a region of integration
			$\Delta_{pp'}^\xi(\theta,\alpha)$ for each type (c) interval
			$(\theta,\alpha)$. Note that not for all $(\theta,\alpha)$ the point
			curve stays above the cutoff (LHS) but there are cases where the
			point curve crosses the cutoff (RHS).}
\label{fig: type c chords}
\end{figure}

The more subtle part is to calculate \eqref{eq: cutoff chord length}
for intervals $(\theta,\alpha)$ of type (c). The difficulty here comes from the
fact that $\Delta_{pp'}$ is bounded by a point curve, not just light rays, as
for the type (b) intervals. This point curve even crosses the cutoff in some
cases, as depicted in \figref{fig: type c chords}.
%--------------------------------------------
%--------------------------------------------

However it is possible to bypass this subtlety for the semicircle by
exploiting several symmetries of the situation.  Since these
symmetries are easiest understood from the bulk point of view, we
choose this picture to explain them. $\Lambda_\xi(\theta,\alpha)$ is,
up to differences due to the regularization scheme, the length of the
segment of the geodesic $(\theta,\alpha)$ that lies inside of
$\Sigma$.
%--------------------------------------------
\begin{figure}
\begin{center}
\includegraphics[scale=0.2]{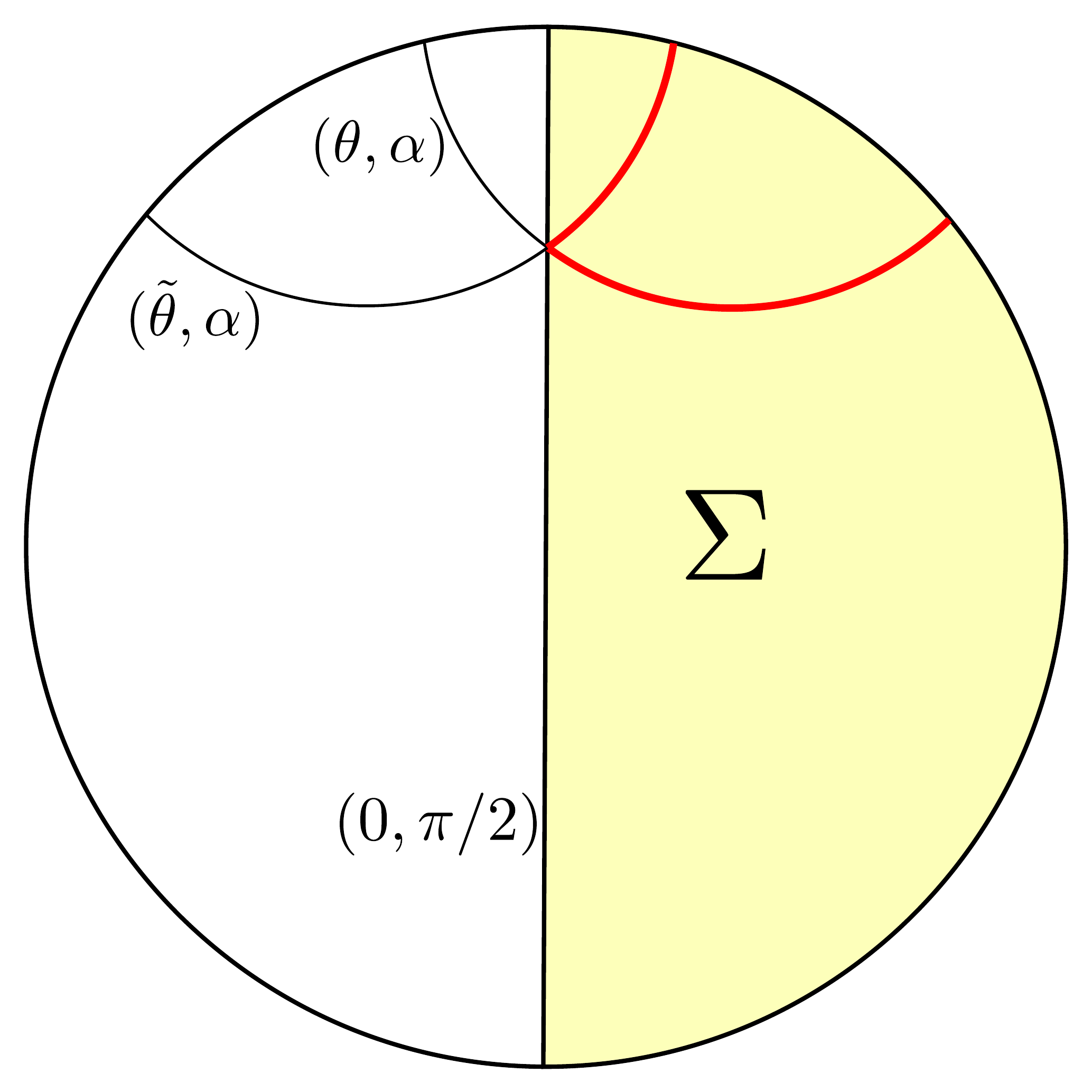} 
\end{center}
\caption{The geodesics $(\theta,\alpha)$ and $(\tilde\theta,\alpha)$ both lie in
			region III in kinematic space. $\Sigma$ is the region below the
			geodesic $(0,\pi/2)$. The chord length of $(\tilde\theta,\alpha)$
			w.r.t. $\Sigma$ is equal to  the length of the part of
			$(\theta,\alpha)$ that lies outside of $\Sigma$. Therefore the
			sum of the chord lengths of $(\theta,\alpha)$ and
			$(\tilde\theta,\alpha)$ gives the total length of $(\theta,\alpha)$.}
\label{fig: counter part geodesics}
\end{figure}
%--------------------------------------------
As depicted in \figref{fig: counter part geodesics} each geodesic
$(\theta,\alpha)$ in III has a counter part $(\tilde\theta,\alpha)$ that also
lies in III, has the same opening angle $\alpha$ and whose chord length is equal
to the length of the part of $(\theta,\alpha)$ that lies outside of
$\Sigma$. Because the cutoff is independent of $\theta$, these chord
lengths together sum up to the length of a
full geodesic. Since we are integrating over all chord lengths to
obtain the complexity, this means we can replace the integration over
the chord lengths of the geodesics in III by an integral over their
full length and multiplying with $1/2$.
Consequently we can replace $\Lambda_\xi$ in the integral over the geodesics
in III in \eqref{eq: calc complexity for semicircle alt} with
\eqref{eq: ent entropy from kin sp} and multiply by $1/2$.
We find
\begin{equation}
\begin{split}
	\cC(\text{semicircle})
	&
	=
	-\frac{6}{\pi c}\Big(
	\int_\xi^{\pi/2}d\alpha\int_{\alpha-\pi/2}^{\alpha+\pi/2}d\theta\left(
		\log\left(\frac{\sin(\alpha)}{\sin(\xi)}\right)
		+
		\xi\cot(\xi)\right)\partial^2_\alpha S(\alpha)
		\Big)
	\\
	&
	=
	2\xi\cot^2(\xi)+2\cot(\xi)+2\xi-\pi
	=
	\frac{4}{\xi}-\pi+\cO(\xi^2)\,.
\end{split}	
\end{equation}

%%%%%%%%%%%%%%%%%%%%%%%%%%%%%%%%%%%%%%%%%%%%%%%%%%%%%%%%%%%%%%%%%%%%%%%%%%%%%%%%
\bibliographystyle{JHEP}
\bibliography{comp-kin-sp}
\end{document}